\documentclass[useAMS,usenatbib]{mn2e}
\usepackage{graphicx}

\title[A semi-automatic procedure for abundance determination of A- and F-type stars]{A semi-automatic procedure for abundance determination of A- and F-type stars}
\author[S. Hekker et al.]{S. Hekker$^{1}$\thanks{e-mail: saskia@oma.be}, Y. Fr\'emat$^{1}$, P. Lampens$^{1}$, P. De Cat$^{1}$, E. Niemczura$^{2}$,
\newauthor O.L. Creevey$^{3}$, J. Zorec$^{4}$\\
$^{1}$Royal Observatory of Belgium, Ringlaan 3, 1180 Brussels, Belgium\\
$^{2}$Astronomical Institute, Wroclaw University, ul. Kopernika 11, 51-622 Wroclaw, Poland\\
$^{3}$Instituto de Astrof\'isica de Canarias, C/ V\'ia Lactea s/n, La Laguna 38205, Tenerife, Spain\\
$^{4}$Institut d'Astrophysique de Paris, UMR7095 CNRS, Universite Pierre $\&$ Marie Curie, 98bis Bd. Arago, 75014 Paris, France}
\begin{document}

\date{Accepted: Received:}

\pagerange{\pageref{firstpage}--\pageref{lastpage}} \pubyear{2009}

\maketitle

\label{firstpage}

\begin{abstract}
A variety of physical processes leading to different types of pulsations and chemical compositions is observed among A- and F-type stars. To investigate the underlying mechanisms responsible for these processes in stars with similar locations in the H-R diagram, an accurate abundance determination is needed, among others. Here, we describe a semi-automatic procedure developed to determine chemical abundances of various elements ranging from helium to mercury for this type of stars. We test our procedure on synthetic spectra, demonstrating that our procedure provides abundances consistent with the input values, even when the stellar parameters are offset by reasonable observational errors. For a fast-rotating star such as Vega, our analysis is consistent with those carried out with other plane-parallel model atmospheres. Simulations show that the offsets from the input abundances increase for stars with low inclination angle of about 4$^{\circ}$. For this inclination angle, we also show that the distribution of the iron abundance found in different regions is bimodal. Furthermore, the effect of rapid rotation can be seen in the peculiar behaviour of the H$\beta$ line.
\end{abstract}


\section{Introduction}
A large variety of phenomena in terms of pulsations and chemical peculiarity is observed among main-sequence A and F stars. The underlying mechanisms of these different phenomena in stars with similar luminosity and temperature are not (yet) well understood. 

The p-mode pulsations in $\delta$ Scuti stars are driven by the kappa (opacity) mechanism active in the second partial ionisation zone of helium. For a long time it has been thought that these pulsations and chemical peculiarity are mutually exclusive \citep[e.g.][]{breger1970}. The existence of chemical peculiarities as well as the suppression of pulsations due to a lack of helium in the partial ionisation zone can be explained by radiative diffusion \citep[e.g.][]{michaud1993}, i.e. gravitational settling of helium and levitation/settling of other elements in layers which are sufficiently stable against turbulent mixing. Notwithstanding this theoretical explanation, stars possessing both chemical peculiarities and pulsations are observed. The co-existence of chemical peculiarity and pulsations can only be explained for more evolved stars, because the temperature range where He II partial ionisation can occur shifts to deeper layers where some He is left \citep{cox1979}. In main-sequence stars the existence of both phenomena can not be fully understood in terms of the radiative diffusion hypothesis. We refer to, e.g., \citet{kurtz2000} for an extensive overview.

$\delta$ Scuti and $\gamma$ Dor pulsators are found in similar regions of the H-R diagram, but the $\gamma$ Dor pulsation periods are an order of magnitude longer. This implies g-mode and not p-mode pulsations. The driving mechanism of the more recently discovered $\gamma$ Dor pulsators is convective blocking near the base of their convective envelopes \citep[e.g.][]{guzik2000,loffler2002,warner2003,dupret2004,dupret2005}. To improve the current models for $\gamma$ Dor stars, better constraints on the fundamental atmospheric parameters including chemical abundances are needed. A detailed abundance analyses for about 20 $\gamma$ Dor stars has been carried out by \citet{bruntt2008}. For this sample they find that the abundance pattern of the $\gamma$ Dor stars is not distinct from constant A- and F-type stars. Similar analyses for more stars is needed to confirm this in a statistical sense.
Only some stars show pulsation characteristics of both $\gamma$ Dor and $\delta$ Scuti type \citep[e.g.][]{henry2005}. Interestingly, most of these, so-called, hybrid stars appear to be metallic line (Am) stars.

Furthermore, some very interesting spectroscopic multiple systems with of A / F components are known, such as DG Leo \citep{fremat2005b} and $\theta ^2$ Tau \citep{lampens2007}. These systems have at least one pulsating component and a non-pulsating or stable one. Further investigation of these systems in terms of chemical composition would allow for a direct study of the relation between multiplicity, pulsations and chemical composition.

Abundance determination is not only very interesting, as emphasised by the science cases described above, but also particularly challenging for A- and F-type stars, because of the occurrence of rapid rotation and/or blended lines in these stars. So far, only a few chemical abundance analysis tools for these stars have been developed by, e.g. \citet{erspamer2002} and \citet{bruntt2002,bruntt2004,bruntt2008} which is based on \citet{valenti1996}.  With the spectra we have obtained for all science cases described above in mind, we independently developed a semi-automatic procedure to determine fundamental stellar parameters (Section 2) and chemical abundances (Section 3), which we test on simulated data (Section 4). The main advantage of such tests is that the input stellar parameters are well known a priory. These tests can provide an estimate of the error due to the procedure, and the uncertainties due to errors in the stellar parameters. Furthermore, gravitational darkening effects induced by fast rotation affect the abundance determination of stars seen nearly pole-on. We analysed Vega (Section 5) to compare results obtained with our method with literature values, and subsequently investigated the effects on abundance determination for simulated fast-rotating stars \citep{fremat2005} seen at several low inclination angles (Section 6). 


\section{Stellar parameter determination \label{stelpar}}
\begin{table}
\centering
\begin{minipage}{\linewidth}
\centering
\caption{Regions fitted to derive the projected rotational velocity}
\label{vsiniregions}
\begin{tabular}{c}
\hline
wavelength range \\
 \AA  \\
 \hline
4260 - 4290\\
4445 - 4460\\
4460 - 4475\\
4485 - 4500\\
\hline
\end{tabular}
\end{minipage}
\end{table}

\subsection{Model atmospheres and flux grids}
The radial velocity (RV), projected rotational velocity ($\upsilon \sin i$), effective temperature (T$_{\rm eff}$) and surface gravity ($\log$ g) are determined from spectra using a procedure partly based on the fitting of high-resolution spectroscopic data with synthetic spectra interpolated in a flux grid. This grid was prepared by applying the SYNSPEC programme \citep[][and references therein]{hubeny1995}, and by adding opacities sources due to Rayleigh scattering and to the H$^{-}$ ions. All calculations were performed with the ATLAS9 LTE atmosphere models computed by \citet{castelli2003}. The microturbulent velocity was fixed at 2 km\,s$^{-1}$, while a solar-type chemical composition \citep{grevesse1998} was considered.

\begin{table}
\centering
\begin{minipage}{\linewidth}
\centering
\caption{Spectral lines used to determine microturbulence for stars of various effective temperatures. 'all' indicates that a spectral line can be used for all considered temperature ranges.}
\label{vmicrolines}
\begin{tabular}{lll}
\hline
wavelength & element & temperature range\\
 \AA  & & K\\
 \hline
4476.0 & Fe~I &  $\leq$ 8250 \\
4508.3 & Fe~II & all\\
4515.5 & Fe~II & all\\
4520.2 & Fe~II & all\\
4558.7 & Cr~II & all\\
4576.5 & Fe~II & $\leq$ 8250\\
4588.2 & Cr~II & $\geq$ 7250\\
4611.3 & Fe~I & $\leq$ 7750\\
4957.5 & Fe~I & $\leq$ 9250\\
5169.0 & Fe~II & $\geq$ 7750\\
5172.7 & Mg~I & all\\
\hline
\end{tabular}
\end{minipage}
\end{table}

\subsection{Procedure}
The procedure we follow for stellar parameter determination (RV, $\upsilon \sin i$, T$_{\rm eff}$,$\log$ g) consists of four consecutive steps \citep[see also][]{fremat2007}. First, RV is determined from the cross-correlation function (CCF) computed with a synthetic spectrum corresponding to the star's known spectral type. Then, $\upsilon \sin i$ is determined based on metallic line fitting in the sensitive regions indicated in Table~\ref{vsiniregions}. The fitting is performed with the Girfit computer code \citep{fremat2006}, where the Minuit Fortran package developed and applied by CERN is used to perform least squares minimisation based on a simplex method \citep{press2002}. All regions in Table~\ref{vsiniregions} are fitted independently with the previously determined RV while $\log$ g, T$_{\rm eff}$ and $\upsilon \sin i$ are free parameters.

\begin{figure*}
\begin{minipage}{0.45\linewidth}
\centering
\includegraphics[width=\linewidth]{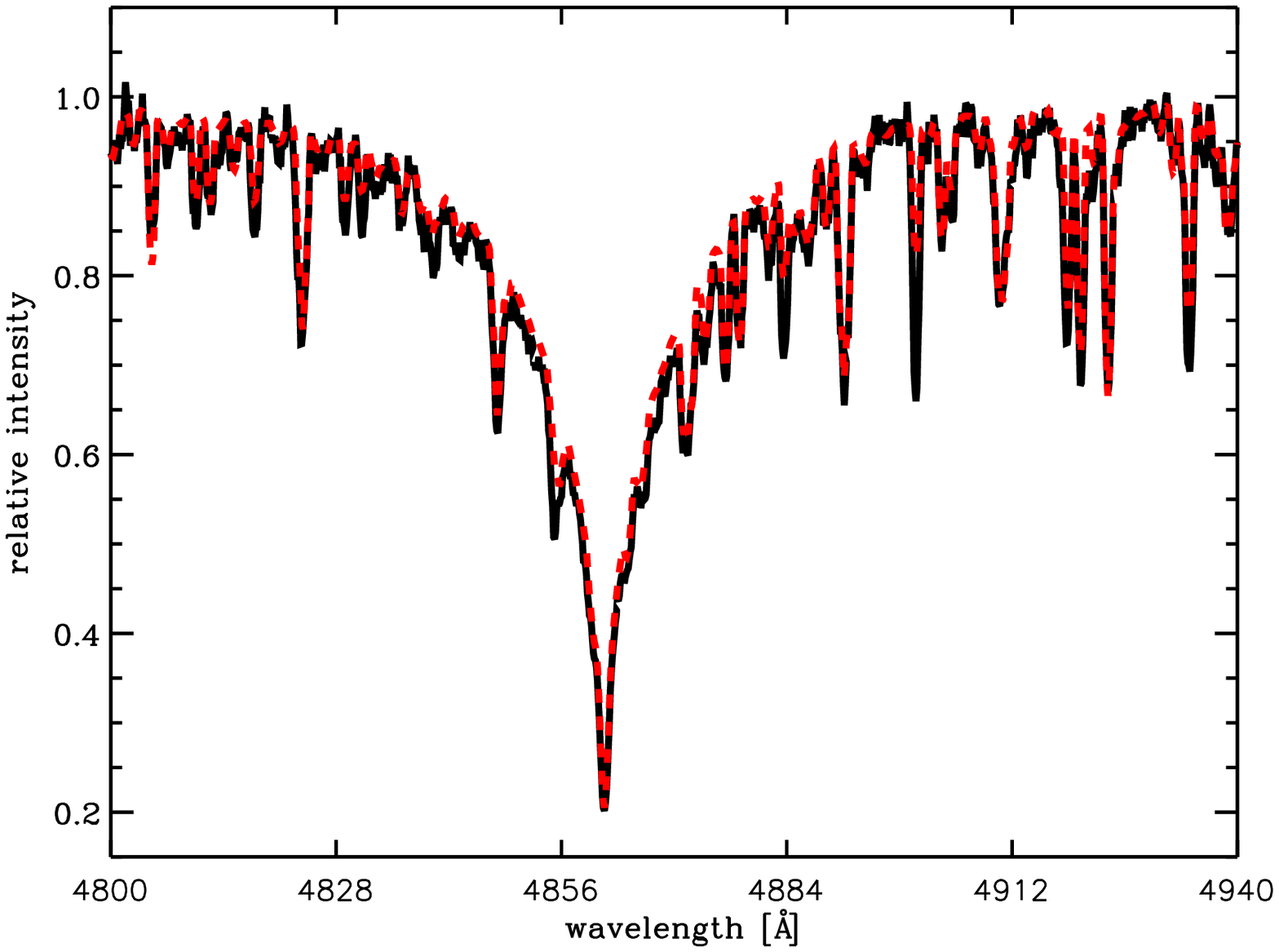}
\end{minipage}
\begin{minipage}{0.45\linewidth}
\centering
\includegraphics[width=\linewidth]{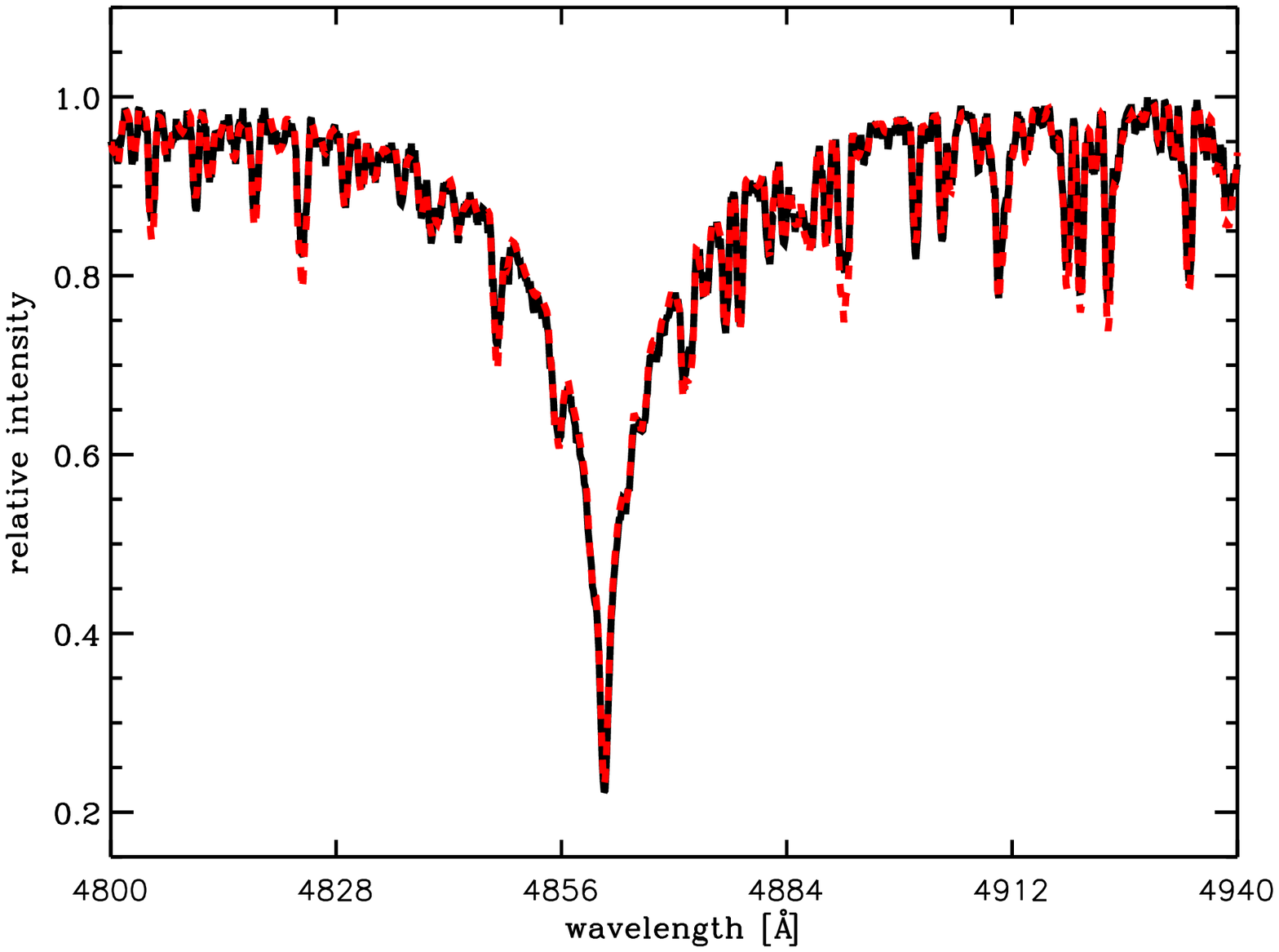}
\end{minipage}
\begin{minipage}{0.45\linewidth}
\centering
\includegraphics[width=\linewidth]{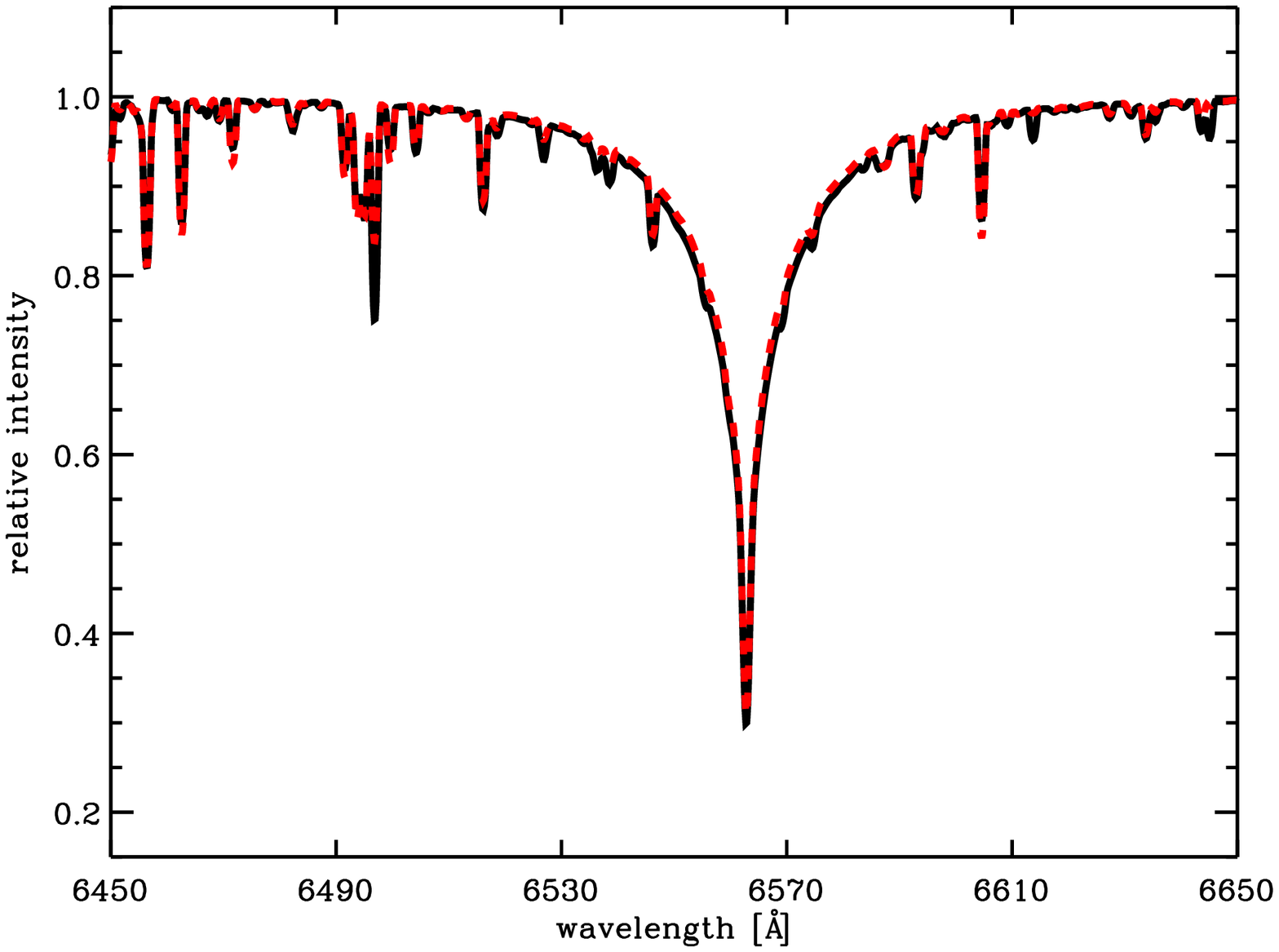}
\end{minipage}
\begin{minipage}{0.45\linewidth}
\centering
\includegraphics[width=\linewidth]{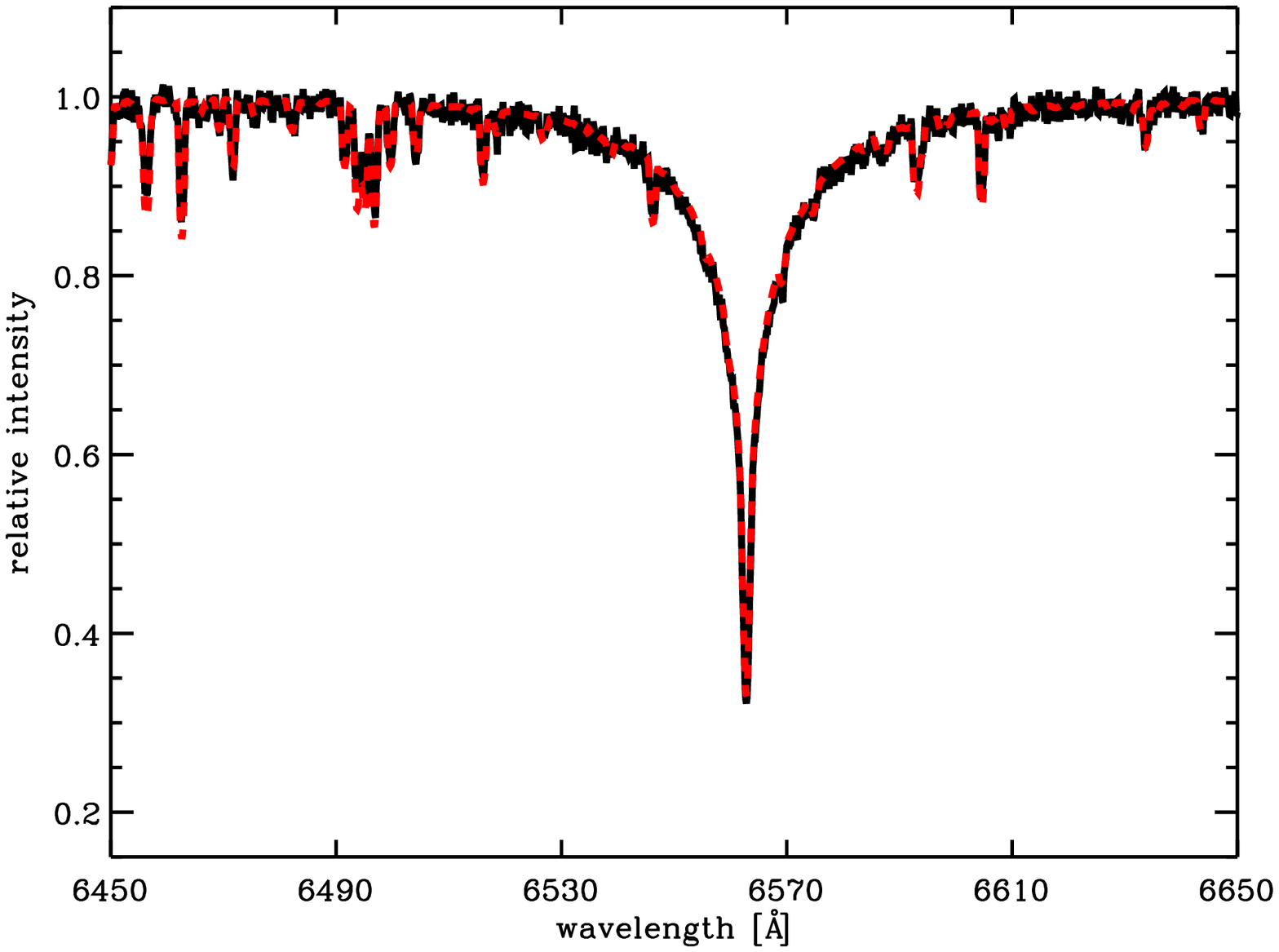}
\end{minipage}
\caption{H$\beta$ (top) and H$\alpha$ (bottom) regions of the synthetic test spectra (black solid line) and the fitted (red dashed line) spectra for star I (left) and star II (right). (Colours only available in the online version.) }
\label{HbHa}
\end{figure*}

The stars we are concerned with are generally cooler than 8500 K, meaning that the hydrogen lines are sensitive to the effective temperature, and insensitive to Stark broadening and surface gravity variations. Therefore, we use Balmer lines to determine T$_{\rm eff}$. For both A- and F-type stars, we use here the H$\beta$ and H$\alpha$ lines, since around these lines we can obtain reliable continuum normalisation. In early A-type stars, a reliable continuum could also be defined around the H$\gamma$ and H$\delta$ lines, but from tests on synthetic spectra we found that this is not the case for cooler stars, which are relatively more affected by metallic line blends. For this reason, and in order to conduct a homogeneous analysis of a sample of A and F stars, we do not take these lines into account. H$\beta$ and H$\alpha$ are fitted with the same packages as described above, with the previously obtained values for $\upsilon \sin i$ and RV, while $\log$ g is set to 4.0 in the first iteration.

Finally, $\log$ g has been derived by combining the Hipparcos parallax, apparent magnitude (reddening is not taken into account at the moment as we currently only have stars of 8$^{th}$ magnitude and brighter at our disposal) and derived T$_{\rm eff}$ to obtain the stars' luminosity \citep{erspamer2003}. T$_{\rm eff}$ and luminosity values are then used to interpolate the stellar mass and radius in the theoretical evolutionary tracks of \citet{schaller1992}, which enables us to estimate the surface gravity. This value of $\log$ g is then adopted in a second iteration step to rederive the effective temperature. 

The results of applying the procedures described above to determine stellar parameters for synthetic test spectra and an observed spectrum of Vega are described in Section 4, 6 and Section 5, respectively.

\begin{table}
\centering
\begin{minipage}{\linewidth}
\centering
\caption{Stellar parameters of the two test spectra. The errors indicated for the computed values are the standard deviations of the values determined in different wavelength regions, systematic errors due to continuum normalisation have to be added. }
\label{stelpartest}
\begin{tabular}{lrrrr}
\hline
 & star I & star I & star II & star II\\
 & input & computed & input & computed\\
 \hline
T$_{\rm eff}$ [K] & 7500 & 7367 $\pm$ 30 & 7000 & 6976 $\pm$ 14\\
$\log$ g (c.g.s) & 3.5 & - & 4.0  & -\\
$\upsilon \sin i$ [km\,s$^{-1}$] & 35.0 & 35.8 $\pm$ 0.8 & 32.1 & 33.2 $\pm$ 0.4\\ 
$\xi_{\rm micro}$ [km\,s$^{-1}$] &  3.70 & 3.7 $\pm$ 0.5 & 1.58 & 2.2 $\pm$ 1.0 \\
\hline
\end{tabular}
\end{minipage}
\end{table}

\section{Abundance analyses \label{abun}}
The abundance determination performed in the present work is based on a spectrum synthesis method and applied to the wavelength region 4500 - 5500 \AA. These limits are set to avoid contamination with telluric lines, which occur at higher wavelength, and continuum normalisation problems. For wavelengths below 4500 \AA~the high number of spectral lines in late A and cooler stars makes it hard to define the continuum for a large wavelength range, which might hamper the abundance analysis. 


Model atmospheres and flux grids described in Section 2.1 with stellar parameters determined with the procedure described in Section 2.2 are used for the present analysis.
Furthermore, the line list provided by \citet{hubeny1995} is used with updated $\log$ gf values from the NIST\footnote{http://physics.nist.gov} database. 

\subsection*{Procedure}

\textbf{1. Microturbulence parameter.} Obtaining microturbulence ($\xi_{micro}$) is an important first step of an accurate abundance determination. We adopt the method described by \citet{gebran2008}. They use the Mg~II triplet at 4480 \AA~and a number of neighbouring unblended weak and moderately strong Fe~II lines to fit for  $\xi_{micro}$ and rotational velocity while only allowing for small variations around the solar value. Here we use different spectral lines, listed in Table~\ref{vmicrolines}, which are considered to be unblended and sensitive to $\xi_{micro}$ in certain temperature intervals. We fit these lines leaving $\xi_{micro}$ as a free parameter and allow only for small changes in the abundances.
The lines mentioned in Table~\ref{vmicrolines} are all fitted separately, which is the approach we use throughout the whole procedure. To check the accuracy of the fit, and the sensitivity to a change in $\xi_{micro}$, the variance and a sensitivity parameter are calculated. This sensitivity parameter is the fractional change in flux of each spectral line calculated for a synthetic fit with $\xi_{micro}$ increased by 2 km\,s$^{-1}$ with respect to the best fit obtained. The 2 km\,s$^{-1}$ level is chosen as an upper limit of the accuracy within which we want to obtain $\xi_{micro}$. The best fits are selected by eye and considering the variance and sensitivity parameter. The mean $\xi_{micro}$ is computed and the standard deviation is used as an estimate of the error.

\textbf{2. Iron.} We proceed by selecting isolated strong iron lines, i.e. lines with an equivalent width larger than 5 m\AA~in a spectrum with the `known' stellar parameters and solar abundance. Then, we search for the closest continuum around these lines in the spectrum and check whether there are no other strong lines in the wavelength interval selected. If this is not the case, the iron line is fitted, while $\xi_{micro}$ can vary within its standard deviation around the mean value found in the previous step.  For the sensitivity parameter of iron we take a 0.1 dex increase in the abundance compared to the best fit. The best fits are selected by eye, while also taking the variance and sensitivity into account. The average value of the fitted $\xi_{micro}$ is kept fixed throughout the remainder of the procedure, while the iron abundance is used as an input value and may vary within twice its standard deviation.

\begin{table}
\centering
\begin{minipage}{\linewidth}
\centering
\caption{Input and computed abundance differences with respect to the solar abundance \citep{grevesse1998} for both star I and II. Nr indicates the number of spectral regions in which an element was fitted.}
\label{testabun}
\begin{tabular}{lrrrrrr}
\hline
el & star I & star I & Nr & star II & star II & Nr\\
  & input & computed &  & input & computed  &\\
 \hline
C & $-$0.20 & 0.05 $\pm$ 0.52 & 15 & $-$0.11 & $-$0.01 $\pm$ 0.17 & 17\\
O & $-$0.19 & & & 0.03 & $-$0.02 $\pm$ 0.20 & 1\\
Na & 0.35 & & & $-$0.12 & &\\
Mg & 0.11 & 0.08 $\pm$ 0.10 & 5 & 0.0 & $-$0.06 $\pm$ 0.10 & 2\\
Al & 0.28 & & & $-$0.21 & &\\
Si & 0.33 & 0.20 $\pm$ 0.15 & 9 & $-$0.09 & $-$0.16 $\pm$ 0.29 & 13\\
P & & & & $-$0.24 & &\\
S & 0.40 & 0.15 $\pm$ 0.23 & 2 &  & 0.08 $\pm$ 0.10 & 2\\
Ca & $-$0.11 & $-$0.16 $\pm$ 0.19 & 7 & $-$0.07 & $-$0.07 $\pm$ 0.12 & 7\\
Sc & $-$0.17 & 0.00 $\pm$ 0.20 & 1 & $-$0.01 & 0.00 $\pm$ 0.10 & 6\\
Ti & $-$0.09 & $-$0.10 $\pm$ 0.10 & 31 & $-$0.13 & $-$0.16 $\pm$0.10 & 33\\
V & 0.60 & 0.50 $\pm$ 0.17 & 2 & 0.06 & 0.04 $\pm$ 0.14 & 6\\
Cr & 0.10 & 0.11 $\pm$ 0.10 & 37 & $-$0.04 & $-$0.05 $\pm$ 0.10 & 57\\
Mn & $-$0.17 & $-$0.03 $\pm$ 0.10 & 3 & $-$0.23 & $-$0.15 $\pm$ 0.10 & 10\\
Fe & 0.13 & 0.04 $\pm$ 0.10  & 93 & $-$0.10 & $-$0.18 $\pm$ 0.10 & 114\\
Co & 0.83 & 0.54 $\pm$ 0.34 & 3 & $-$0.01 & 0.06 $\pm$ 0.10 & 15\\
Ni & 0.37 & 0.34 $\pm$ 0.10 & 36 & $-$0.16 & $-$0.15 $\pm$ 0.10 & 43\\
Cu & 0.32 & 0.30 $\pm$ 0.20 & 1 & $-$0.69 & $-$0.80 $\pm$ 0.13 & 2\\
Zn & 0.35 & 0.53 $\pm$ 0.20 & 1 & $-$0.35 & &\\
Sr & 0.82 & & & $-$0.07 & &\\
Y & 0.95 & 0.92 $\pm$ 0.10 & 4 & $-$0.01 & 0.04 $\pm$ 0.27 & 5\\
Zr & 0.63 & 0.75 $\pm$ 0.20 & 1 & $-$0.27 & &\\
Ba & 1.18 & 0.98 $\pm$ 0.27 & 2 & 0.39 & &\\
La & 0.94 & 0.86 $\pm$ 0.20 & 1 & 0.20 & 0.12 $\pm$ 0.20 & 1\\
Ce & 0.92 & & & 0.11 & 0.01 $\pm$ 0.10 & 2\\
Pr & & & & 0.40 & &\\
Nd & 0.71 & & & $-$0.01 & 0.08 $\pm$ 0.13 & 4\\
Sm & 0.83 & & & 0.08 & &\\
Eu & 1.15 & & & $-$0.10 & &\\
\hline
\end{tabular}
\end{minipage}
\end{table}

\textbf{3. Iron peak elements.} We next fit the isolated iron peak elements, i.e. Ti, V, Cr, Mn, Co, Ni and Cu. We fit spectral regions where lines of these elements and possibly iron lines are strong. The variance and sensitivity parameter are calculated, and the best fitted spectral regions are selected. The mean values of the abundances found for each element are used as input for the abundance computation in spectral regions which contain only strong iron peak elements. From this analysis the abundance values for the iron peak elements are determined, which in turn are used to compute abundances for as many elements as possible ranging from helium to mercury.

\begin{figure*}
\begin{minipage}{0.45\linewidth}
\centering
\includegraphics[width=\linewidth]{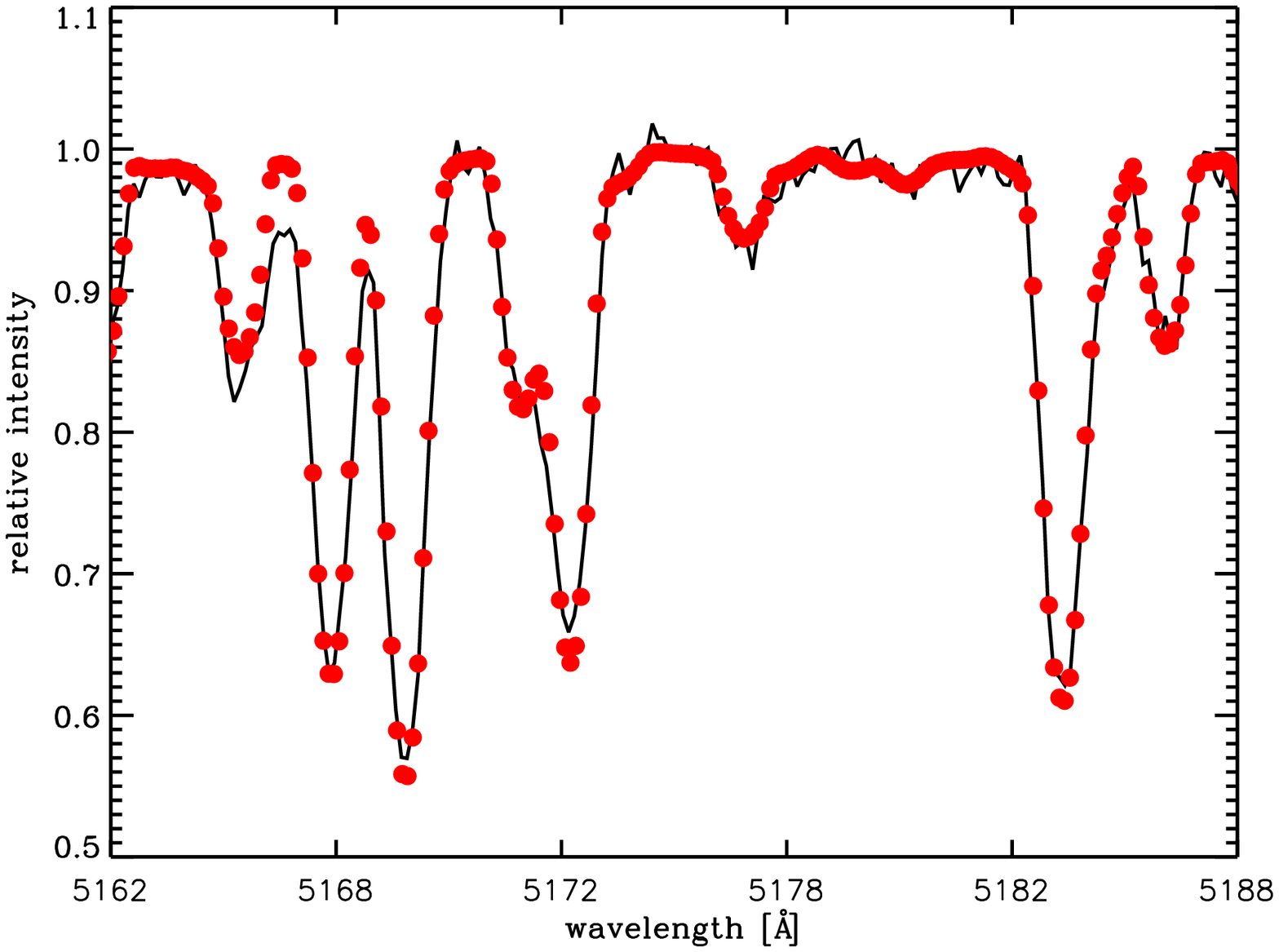}
\end{minipage}
\begin{minipage}{0.45\linewidth}
\centering
\includegraphics[width=\linewidth]{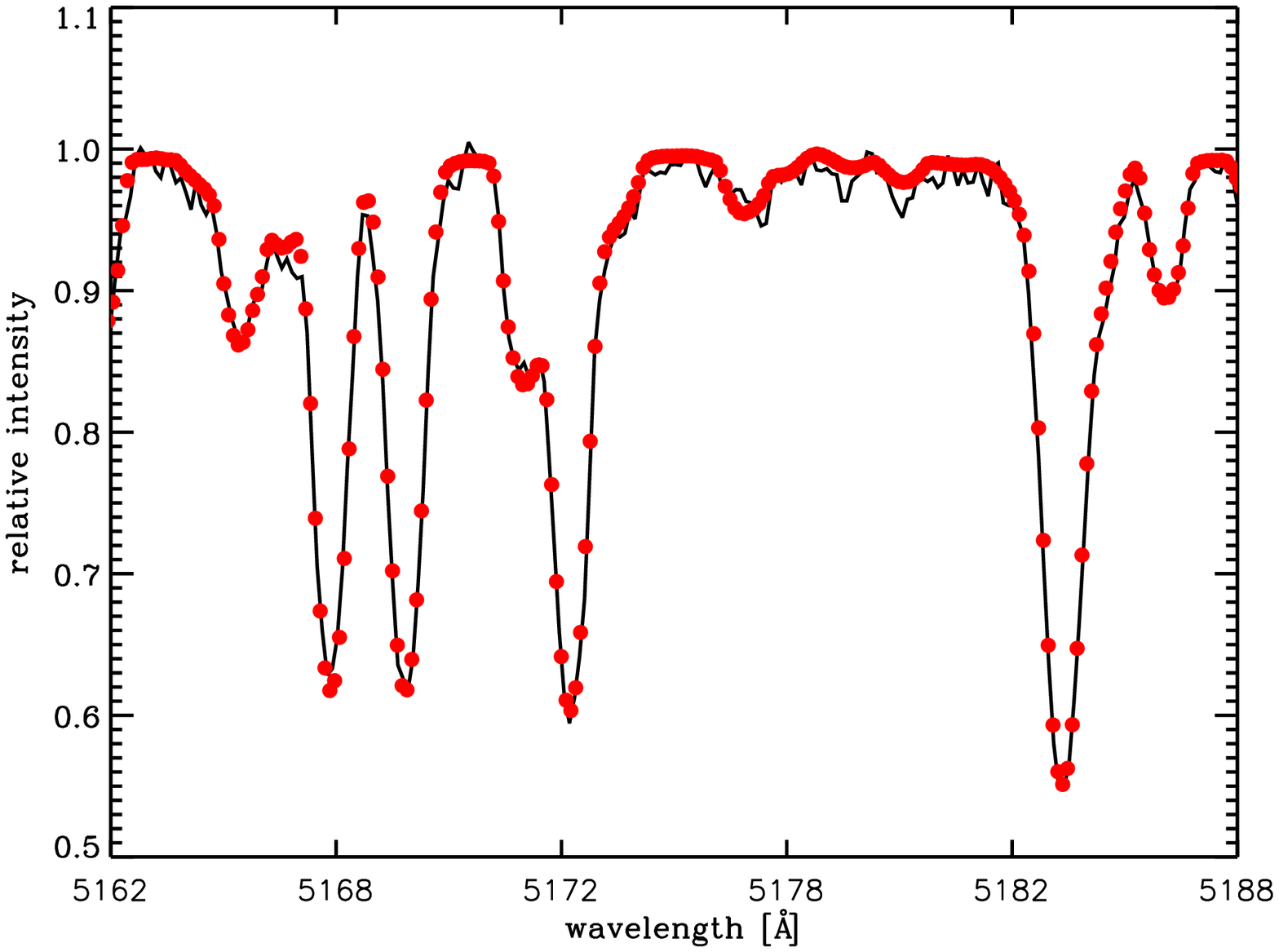}
\end{minipage}
\begin{minipage}{0.45\linewidth}
\centering
\includegraphics[width=\linewidth]{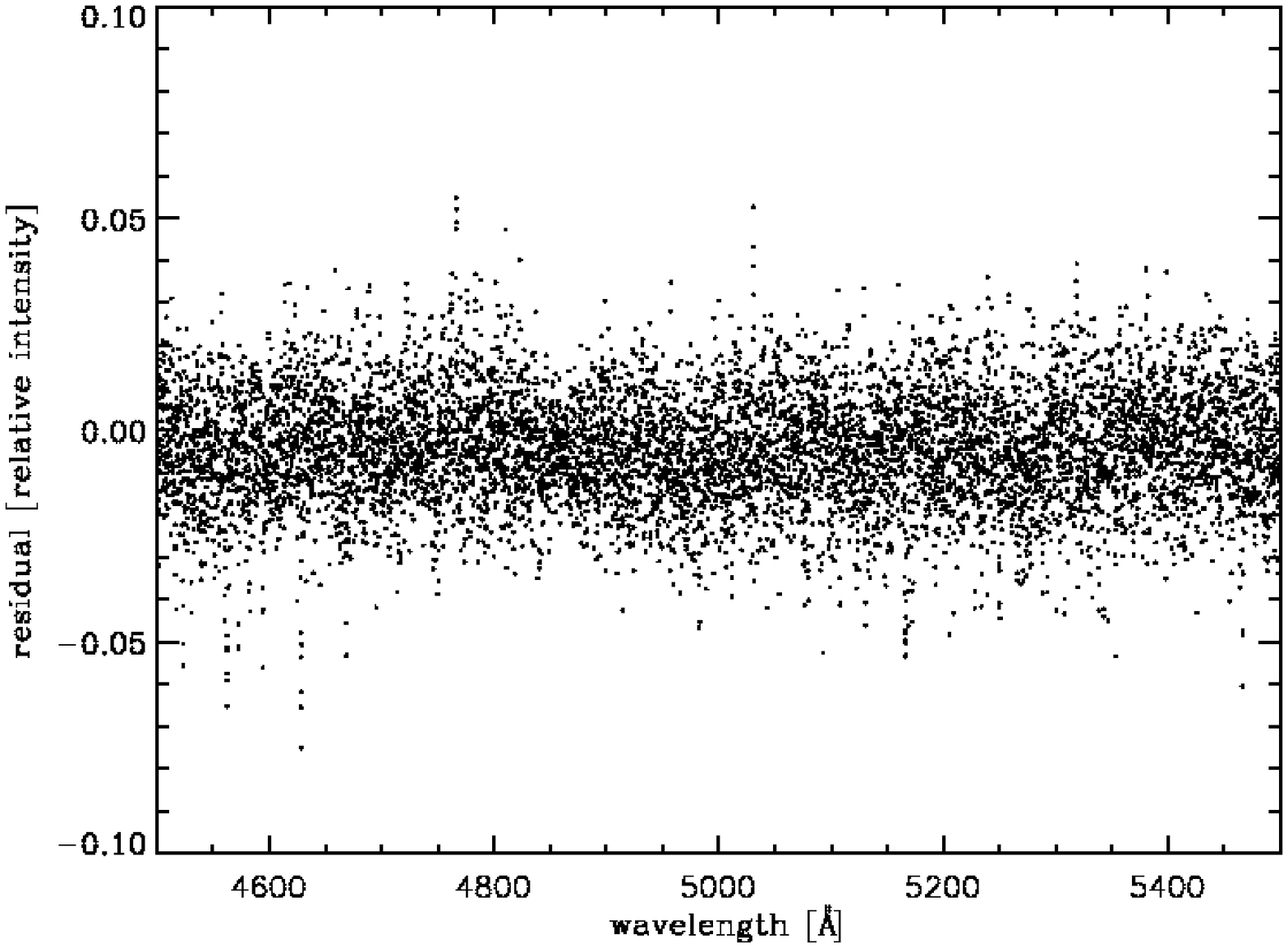}
\end{minipage}
\begin{minipage}{0.45\linewidth}
\centering
\includegraphics[width=\linewidth]{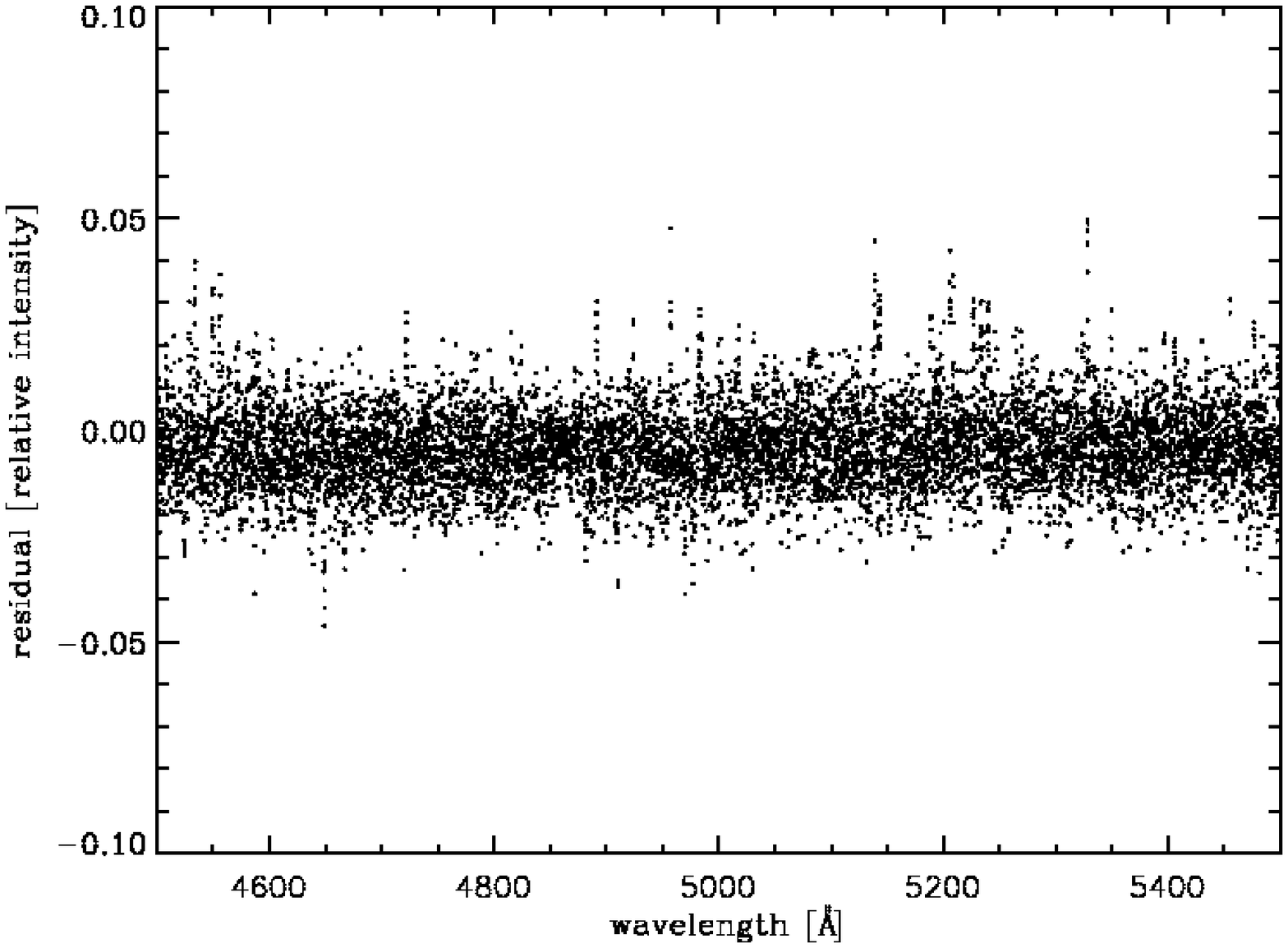}
\end{minipage}
\begin{minipage}{0.45\linewidth}
\centering
\includegraphics[width=\linewidth]{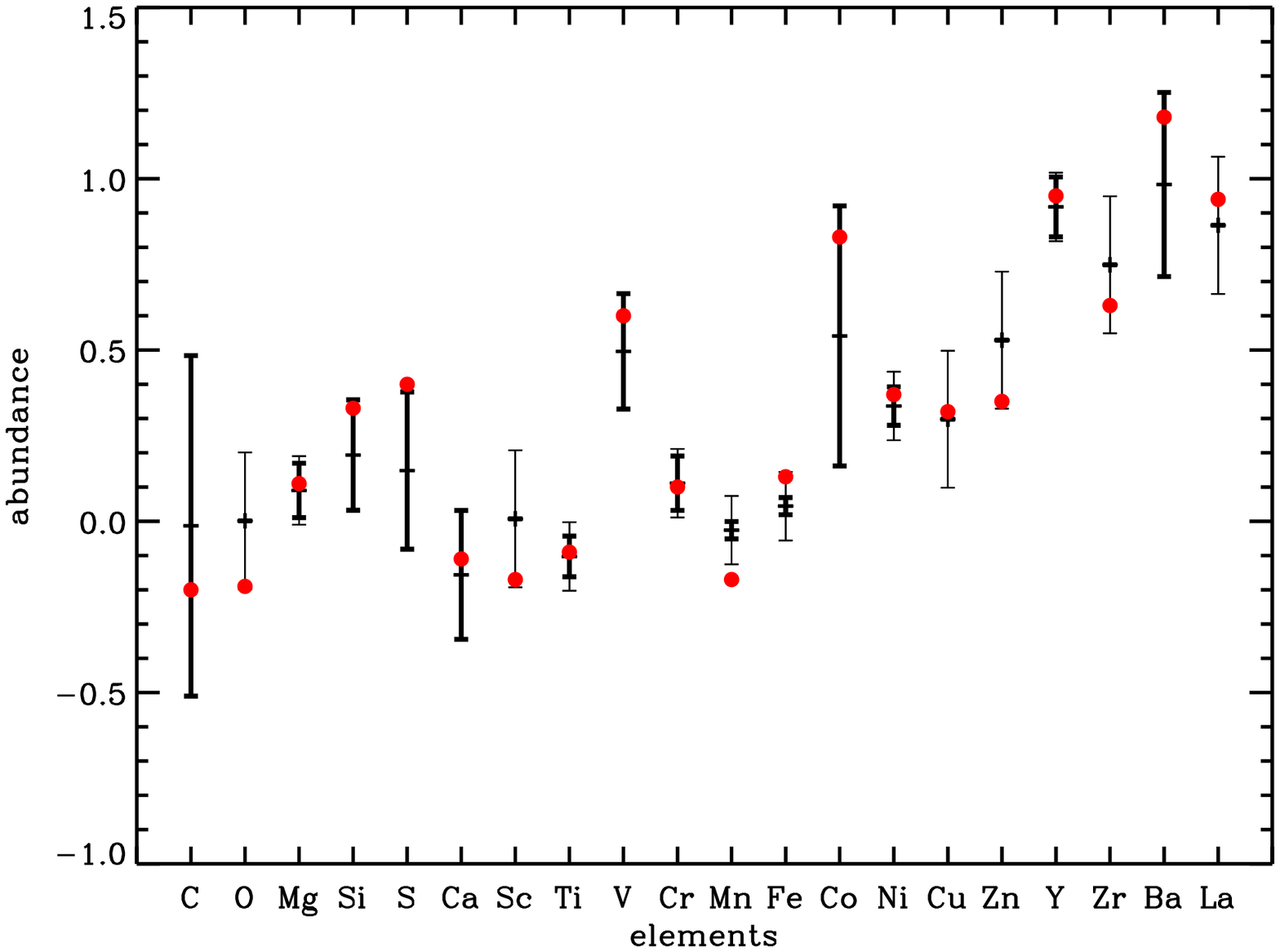}
\end{minipage}
\begin{minipage}{0.45\linewidth}
\centering
\includegraphics[width=\linewidth]{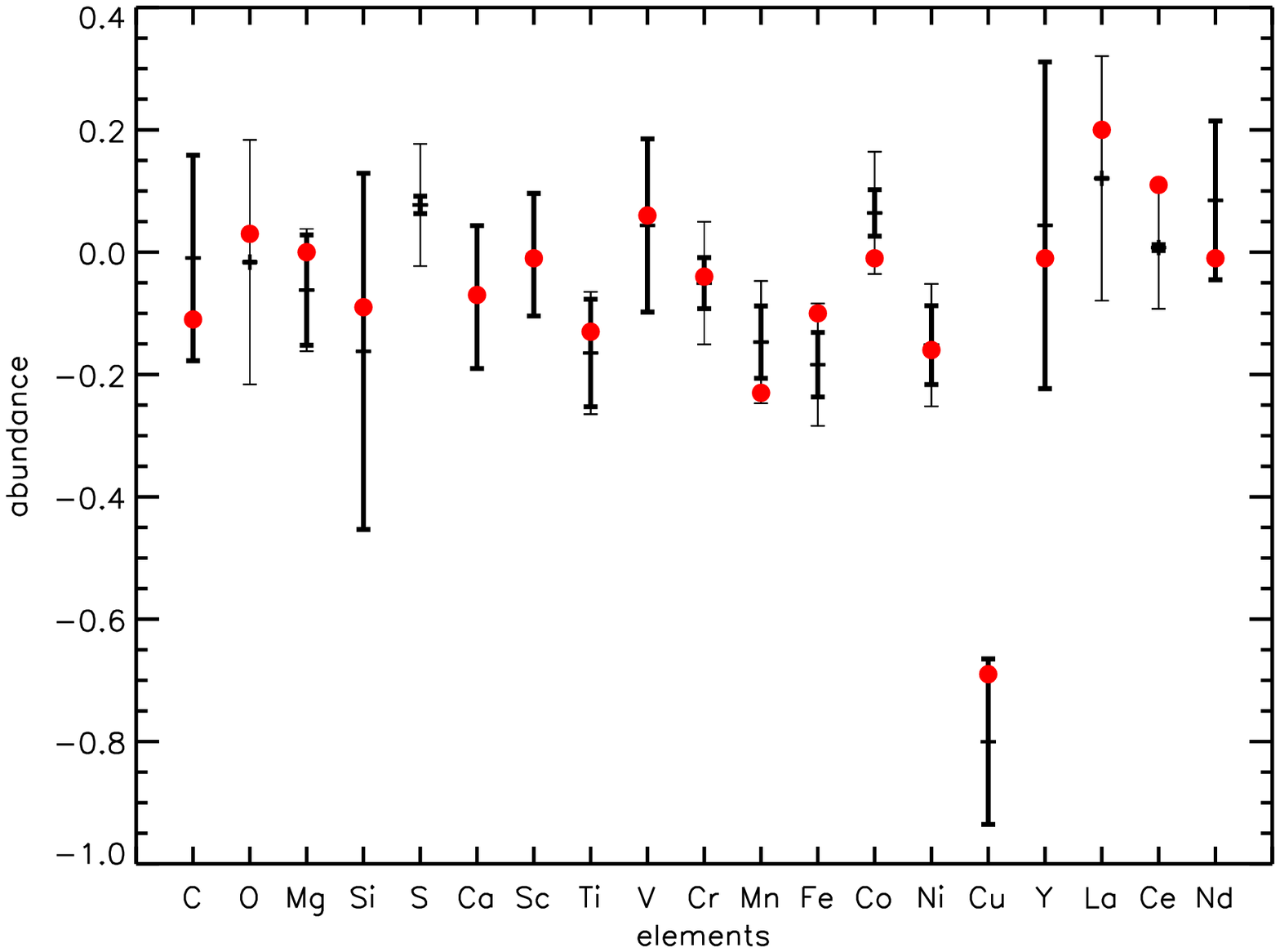}
\end{minipage}
\caption{{\sl Top:} Region of input spectrum (black solid line) and fitted (red dots) spectrum for star I (left) and star II (right). {\sl Middle:} Residuals between input and fitted spectra. {\sl Bottom:} Input abundance (red) and computed abundances (black) of the individual elements determined using the method described in Section 3 (colours only in the online version). The thick error bars are the standard deviations, which serve as the error when they are larger than 0.1. The thin error bars are the inferred errors of 0.1 for an element fitted at several spectral ranges and 0.2 for an element fitted in only one spectral region.}
\label{abunstar}
\end{figure*}

\textbf{4. Final abundances.} For the final abundance determination, we consider all spectral lines with equivalent width $\geq 2$ m\AA~ present in a spectrum with solar abundance, and a continuum is sought around these. All lines stronger than 2 m\AA~in the selected region are fitted, while the earlier determined $\xi_{micro}$ is kept fixed. The abundances of the iron peak elements obtained in the previous step are used (if present) as input, but these values can be changed within twice their standard deviation. For each element fitted in a certain region, the sensitivity parameter, as well as the variance, are computed. The worst fits are excluded and the mean abundances of all fitted elements are computed using the sensitivity parameter as weight. The standard deviation is used as an estimate of the error. We set the error to 0.1 dex when the standard deviation is smaller than 0.1. For elements for which we only fit one spectral line the error is taken to be 0.2 dex.

\begin{figure*}
\begin{minipage}{0.45\linewidth}
\centering
\includegraphics[width=\linewidth]{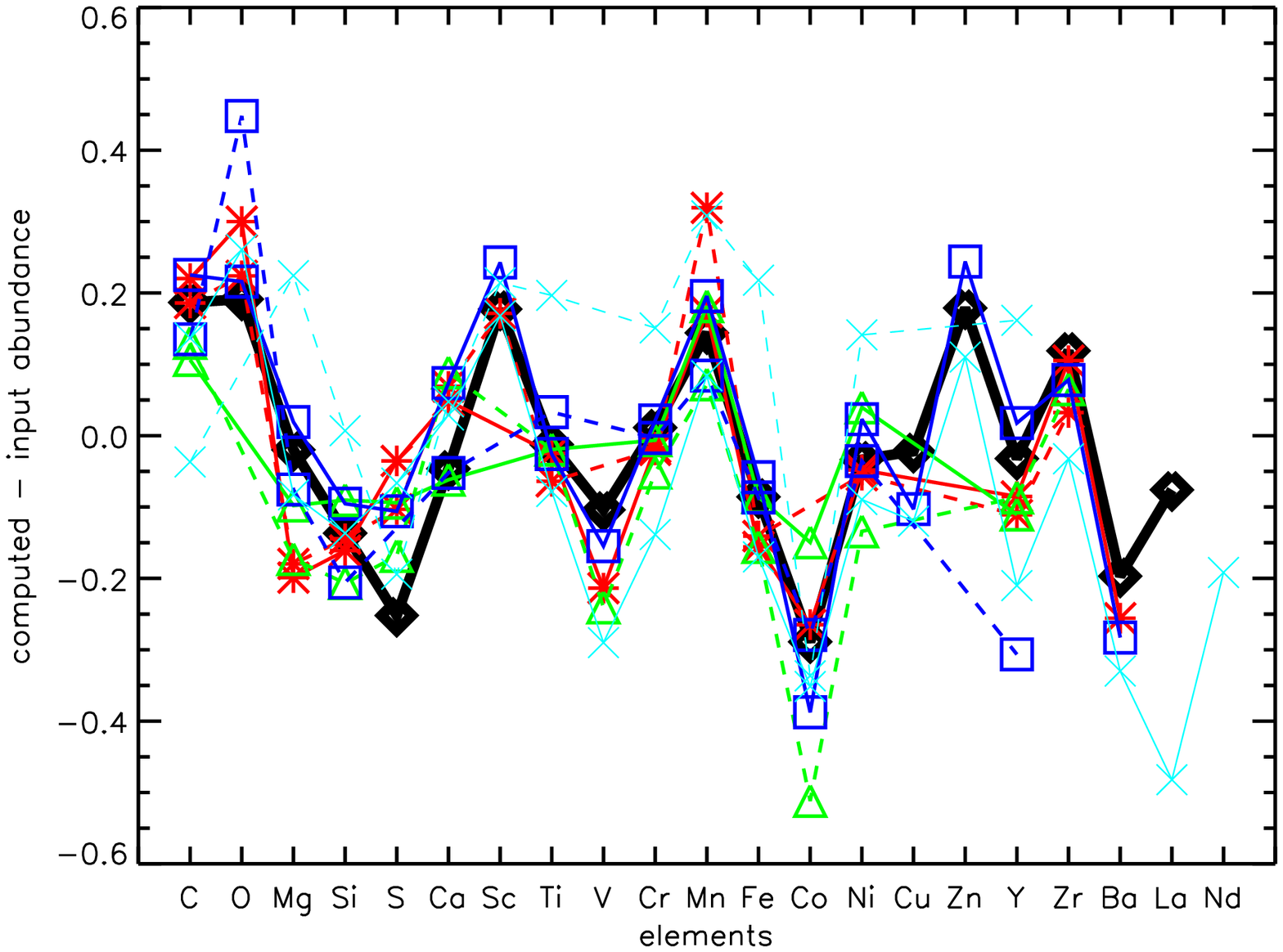}
\end{minipage}
\begin{minipage}{0.45\linewidth}
\centering
\includegraphics[width=\linewidth]{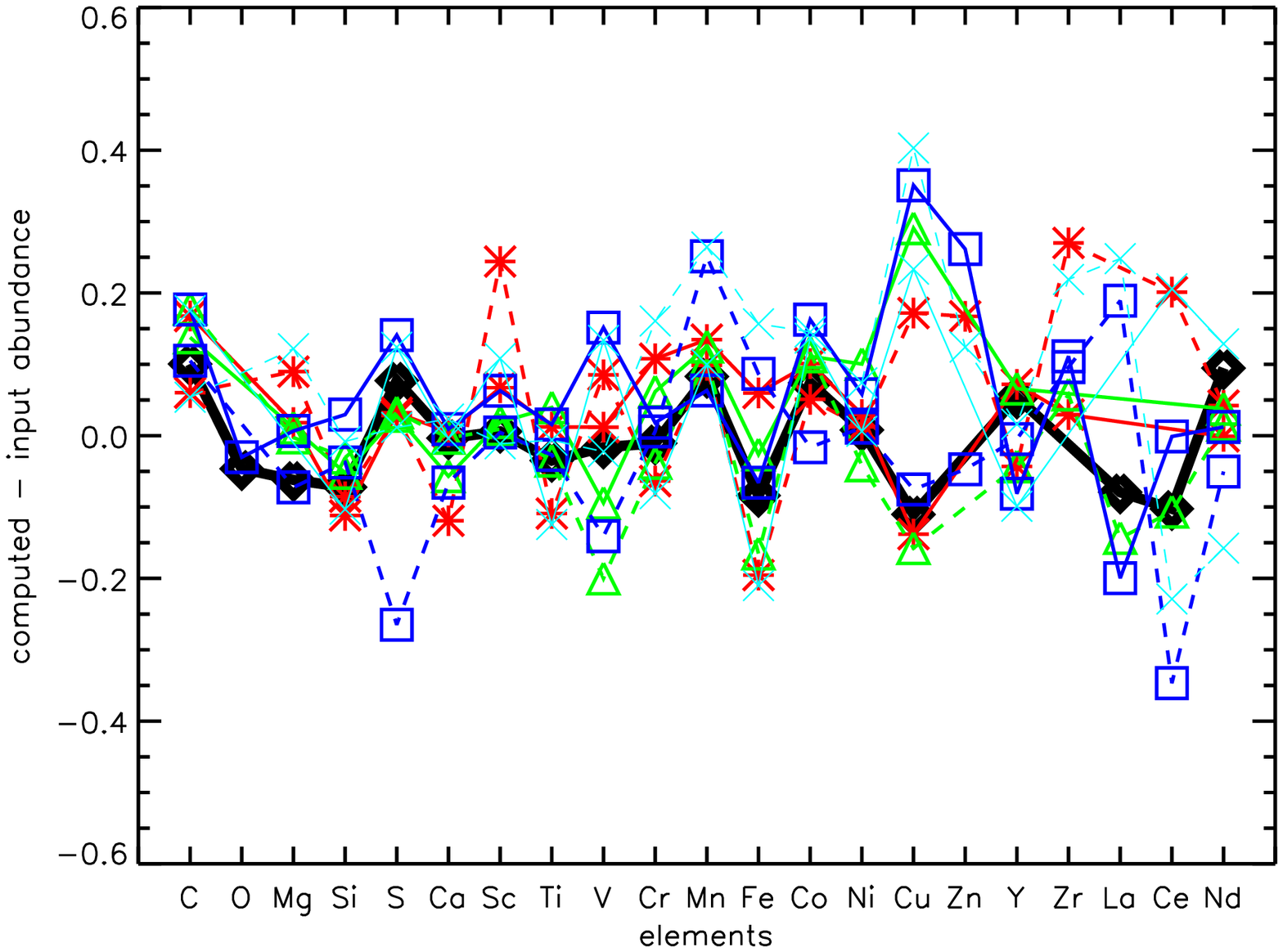}
\end{minipage}
\begin{minipage}{0.45\linewidth}
\centering
\includegraphics[width=\linewidth]{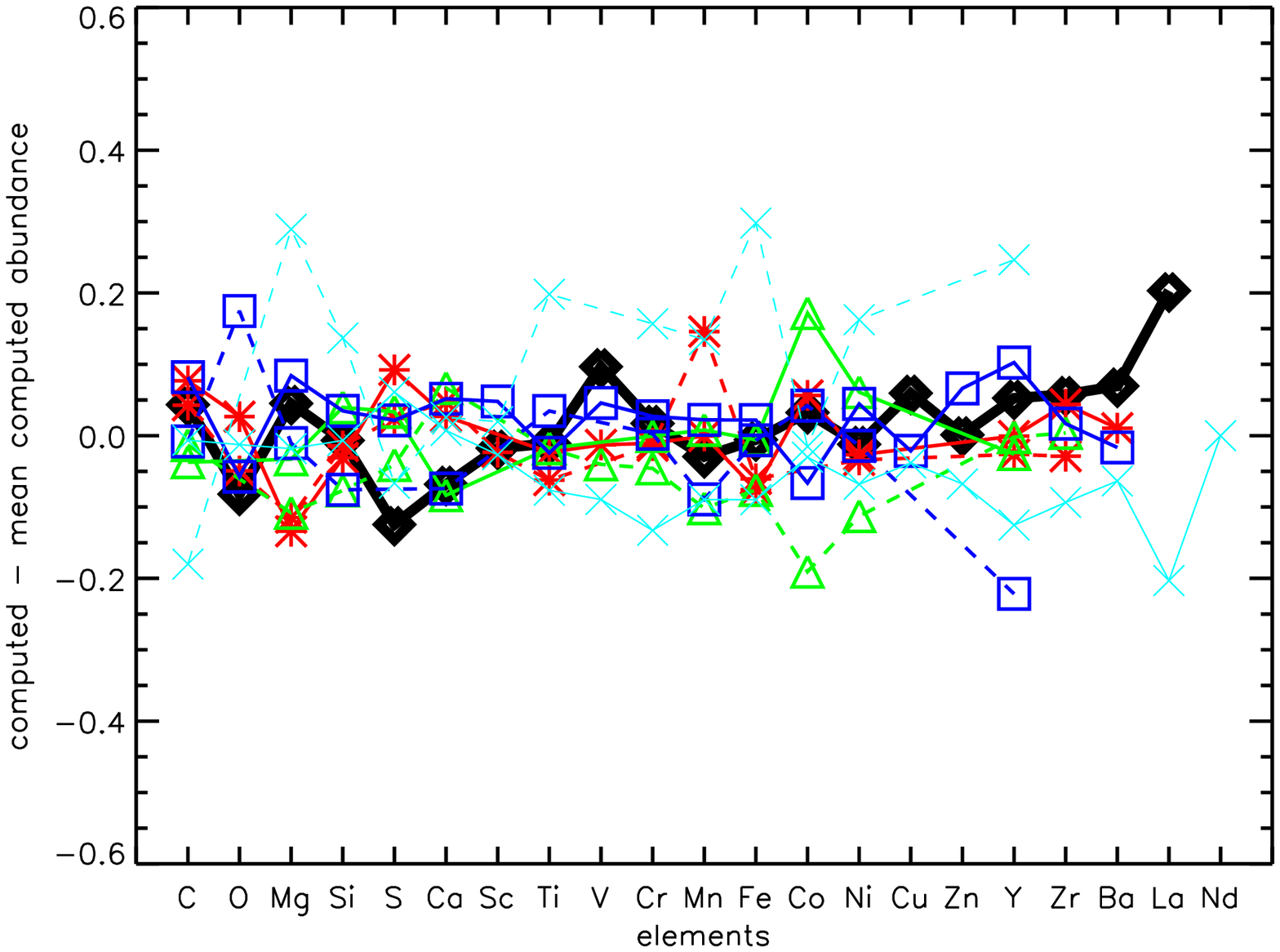}
\end{minipage}
\begin{minipage}{0.45\linewidth}
\centering
\includegraphics[width=\linewidth]{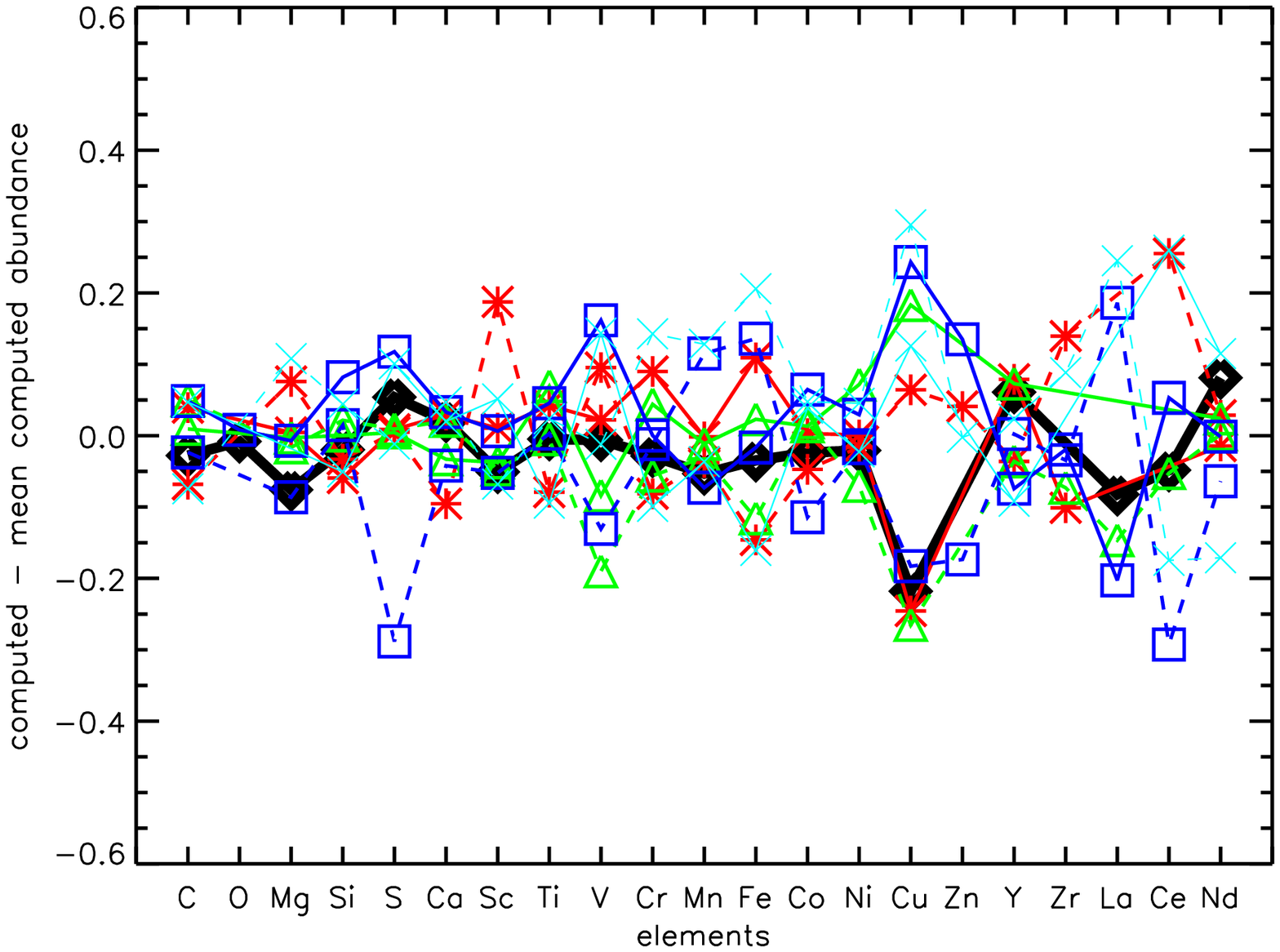}
\end{minipage}
\caption{The computed abundances for different stellar parameters minus the input abundances (top) and minus the mean computed abundances (bottom) for star I (left) and star II (right). The black diamonds (solid black line) indicate the results of the computation with the correct stellar parameters of the synthetic test spectra. The coloured symbols are results from computations with different stellar parameters: red asterisks, green triangles, blue squares, light blue crosses for increased / decreased $\log$ g, T$_{\rm eff}$, $v \sin i$ and $\xi_{micro}$ respectively (colours only in the online version). The solid (dashed) lines connect values computed with increased (decreased) values of the stellar parameters with respect to the input values. The lines are only plotted for visual purposes.}
\label{starres}
\end{figure*}

A last check is performed by inspecting the residuals between the observed and fitted spectra. In some cases, the residuals are larger for specific elements. The corresponding values are adjusted, to reduce the residuals and thus improve the final abundance determination.

\section{Results for synthetic test spectra \label{ressyn}}

To test the procedures described in the previous sections, we computed two synthetic test spectra with a signal to noise ratio of $\approx$80 and a resolving power of $\approx$40\,000. For both spectra we determined the stellar parameters and abundances. The input and computed stellar parameters of the synthetic stars are given in Table~\ref{stelpartest}. As described in Section 2.2, the surface gravities are computed from theoretical stellar evolutionary tracks, using the apparent magnitude and parallax. As the simulation and re-calculation would imply an exact inverse operation, we omitted such procedure here and only computed $\upsilon \sin i$ and T$_{\rm eff}$, by taking into account the input $\log$ g value. We checked the influence of an offset in $\log$ g on the determination of T$_{\rm eff}$ by increasing / decreasing $\log$ g with 0.2. This induced changes in T$_{\rm eff}$ of less than twice the errors quoted in Table~\ref{stelpartest}. Figure~\ref{HbHa} shows the H$\beta$ and H$\alpha$ regions of the synthetic test spectra, together with the best fits we obtained for the stellar parameters.

For the abundance analysis test, we used the input stellar parameters and the same atomic line list with which the spectra are computed. So, no errors due to stellar parameters or inaccurate values in the atomic line list can occur (mainly errors in oscillator strength ($\log$ gf)). In this way we can really test the sensitivity of the procedure. In a second step (see below) the influence of offsets in the stellar parameters will be closely examined.

For $\xi_{micro}$, we found 3.7 $\pm$ 0.5 km\,s$^{-1}$ for star I and 2.2 $\pm$ 1.0 km\,s$^{-1}$ for star II, which is consistent with the input values provided in Table~\ref{stelpartest}. 
Results of the abundance analyses are listed in Table~\ref{testabun} and shown in Fig.~\ref{abunstar}, together with the input and residual spectra. Throughout this work, the mentioned abundance values are always relative abundances with respect to the solar value \citep{grevesse1998}. For Ti, Cr, Fe and Ni most absorption lines with considerable strength are present and these elements are most important for the overall metallicity of the star. Apart from Mn and S for star I, abundances of all elements we computed are consistent, within their error-bars, with the input values. The mean difference in normalised flux between the input and fitted spectrum is -0.004 (input -- fitted spectrum) for both stars with a standard deviation of 0.013 and 0.010 for star I and II, respectively.

\begin{table}
\centering
\begin{minipage}{\linewidth}
\centering
\caption{$\xi_{micro}$ for the tests with offsets in effective temperature, surface gravity and rotational velocity. The input values for $\xi_{micro}$ are 3.7 and 1.58 km\,s$^{-1}$ for test star I and II, respectively.}
\label{vmicro}
\begin{tabular}{lrr}
\hline
 & star I & star II\\
 & $\xi_{micro}$  [km\,s$^{-1}$]  & $\xi_{micro}$  [km\,s$^{-1}$] \\
 \hline
T$_{\rm eff}$ + 100 [K]  & 4.2 & 2.1\\
T$_{\rm eff}$ $-$ 100 [K]  & 3.8 & 2.3\\
$\log$ g + 0.1 (c.g.s) & 4.3 & 1.3\\
$\log$ g $-$ 0.1 (c.g.s)  & 4.2 & 2.9\\
$v \sin i$ + 5 [km\,s$^{-1}$] & 4.3 & 2.7\\ 
$v \sin i$ $-$ 5 [km\,s$^{-1}$] & 2.8 & 0.2\\ 
\hline
\end{tabular}
\end{minipage}
\end{table}

These test spectra are also suitable to investigate the influence of an offset in effective temperature, surface gravity and rotational velocity. Therefore, we computed $\xi_{micro}$ and abundances with a temperature of $\pm$ 100 K, leaving the surface gravity and rotational velocity at the correct value. Similar tests are performed for an increase  / decrease of 0.1 in surface gravity and 5 km\,s$^{-1}$ in rotational velocity. 

The computed $\xi_{micro}$ for the different tests are given in Table~\ref{vmicro}. The changes in $\xi_{micro}$ due to changes in stellar parameters are of the order of the error found for $\xi_{micro}$ when we use the correct stellar parameters. Only for a decrease in $v \sin i$ the value for $\xi_{micro}$ decreases by about twice the error. 
In order to test the influence of an offset in $\xi_{micro}$, the abundances are recomputed with $\xi_{micro}$ $\pm$ 2 km\,s$^{-1}$. The resulting abundances are shown in Fig.~\ref{starres}.

When comparing the results of the tests described above, it becomes clear that the scatter of the computed abundance values around the input abundance of the synthetic test spectra is dominated by scatter inherent to the method (see top panels of Fig.~\ref{starres}). The standard deviations of all computed abundances with respect to the input abundances are 0.17 dex and 0.12 dex for star I and II, respectively. The standard deviations around the mean of the computed abundances for the different stellar parameters are 0.085 dex and 0.10 dex for the respective stars (see bottom panels of Fig.~\ref{starres}). From these tests we conclude that the accuracy of the computed abundances is limited by the procedure itself and to a lesser extent by an inaccurate knowledge of the stellar parameters when uncertainties in these are equal to 100 K (T$_{\rm eff}$), 0.1 dex ($\log$ g), 5 km\,s$^{-1}$ ($v \sin i$) and 2 km\,s$^{-1}$ ($\xi_{micro}$) and when inaccuracies in the atomic line list can be neglected. Note that underestimating $v \sin i$ has the largest influence on $\xi_{micro}$ and the abundances as seen from the light blue crosses connected with the dashed line in Fig.~\ref{starres}.

\section{Results for Vega \label{resvega}}
\begin{figure}
\begin{minipage}{\linewidth}
\centering
\includegraphics[width=\linewidth]{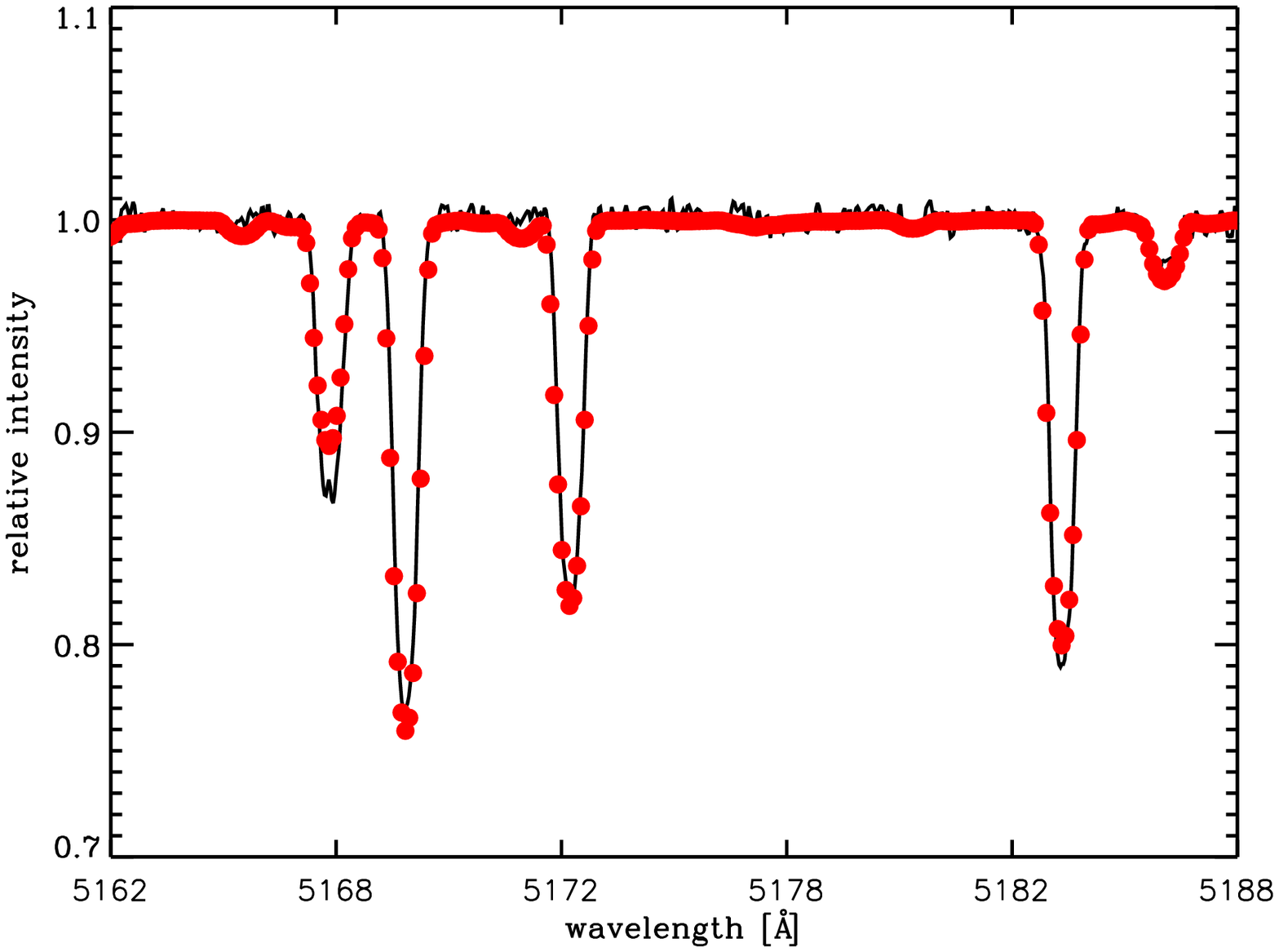}
\end{minipage}
\begin{minipage}{\linewidth}
\centering
\includegraphics[width=\linewidth]{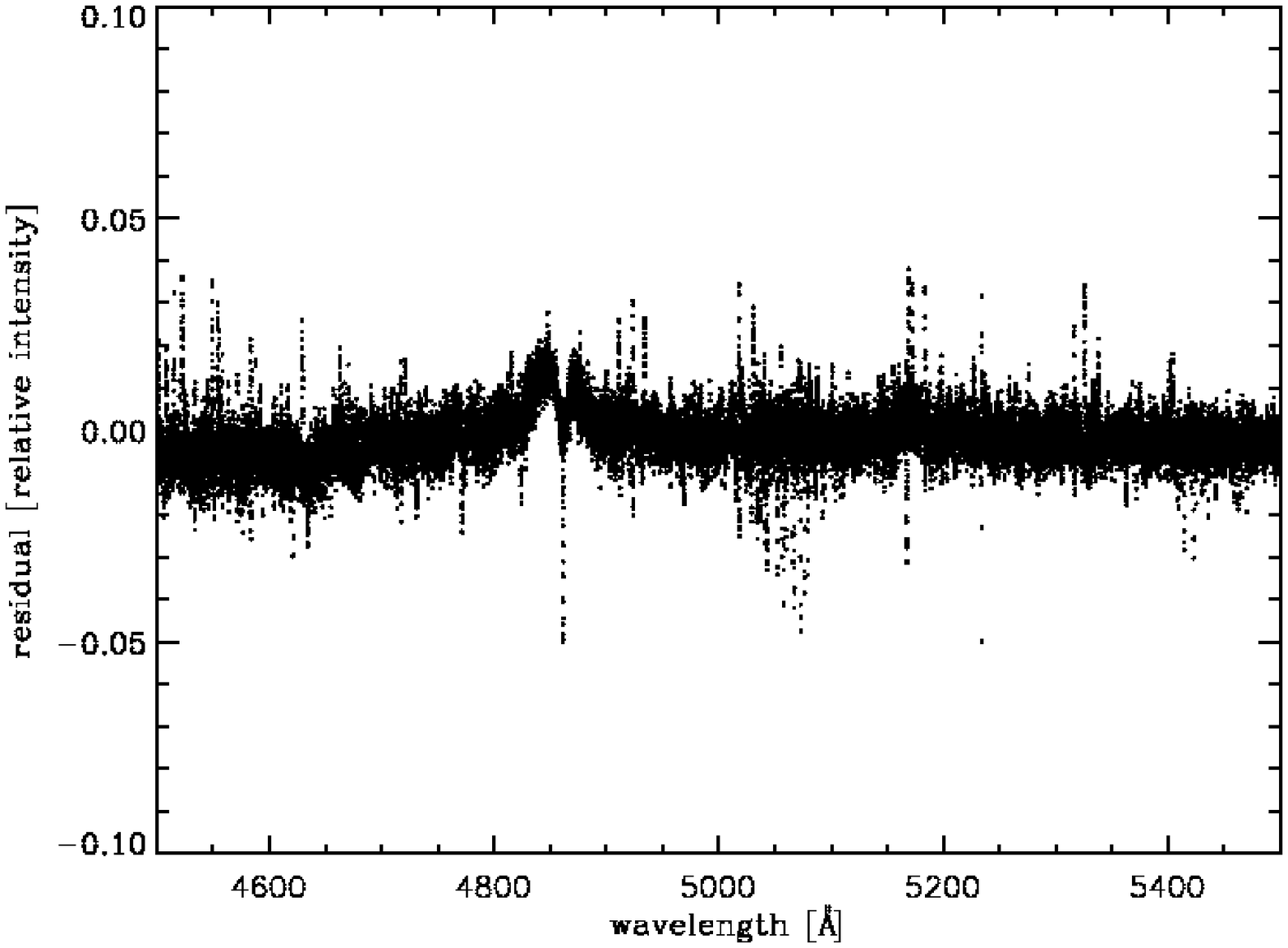}
\end{minipage}
\caption{Observed (black solid line) and fitted (red dots) spectra for Vega (top) with residuals (bottom) (colours only in the online version).}
\label{vega}
\end{figure}

\begin{figure}
\centering
\includegraphics[width=\linewidth]{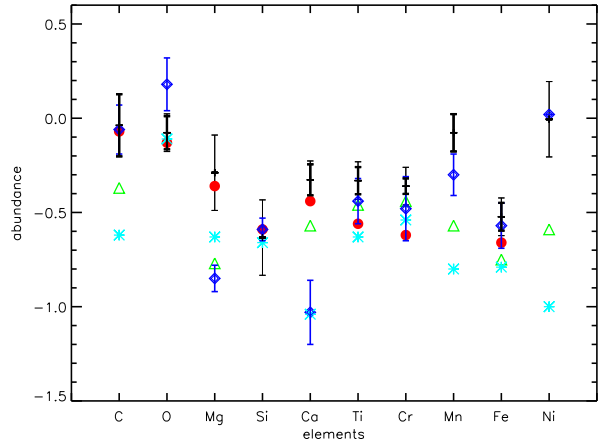}
\caption{Abundances of Vega determined using the method described in this paper (black). The coloured symbols are results from the literature: red dots, green triangles, blue diamonds and light  blue asterisks are results from \citet{erspamer2002}, \citet{adelman1990}, \citet{qiu2001} and \citet{yoon2008}, respectively (colours only in the online version).}
\label{abunvega}
\end{figure}

As a subsequent test we analysed a spectrum of Vega (A0 star with apparent magnitude of 0.03 mag in V) from the ELODIE database. Vega is often used as a reference star, although this is a fast-rotating star seen nearly pole-on, which influences the shape of the spectral lines. It is a bright star for which high signal-to-noise ratio spectra are available and for which several abundance analyses are already performed by other groups, providing comparison data for this test. The stellar parameters determined using our procedure are: T$_{\rm eff}$ = 9560 $\pm$ 170 K, $\log$ g = 4.0 $\pm$ 0.3, $v \sin i$ = 25.4 $\pm$ 2.3 km\,s$^{-1}$ and the radial velocity = -14.0 $\pm$ 1.0 km\,s$^{-1}$, in good agreement with those determined from previous works \citep[e.g.][]{erspamercat2002}. We obtained a $\xi_{micro}$ of 2.4 $\pm$ 0.6 km\,s$^{-1}$, and show a region of the best fitted spectrum, together with residuals in Fig.~\ref{vega}.
The mean difference between the observed and fitted spectrum of Vega is -0.0014 (observed -- fitted spectrum in relative intensity) with a standard deviation of 0.007. The large feature at about 4860 \AA, H$\beta$, could be due to deformations in the line caused by fast rotation, as explained in Section 6.

\begin{figure*}
\begin{minipage}{0.45\linewidth}
\centering
\includegraphics[width=\linewidth]{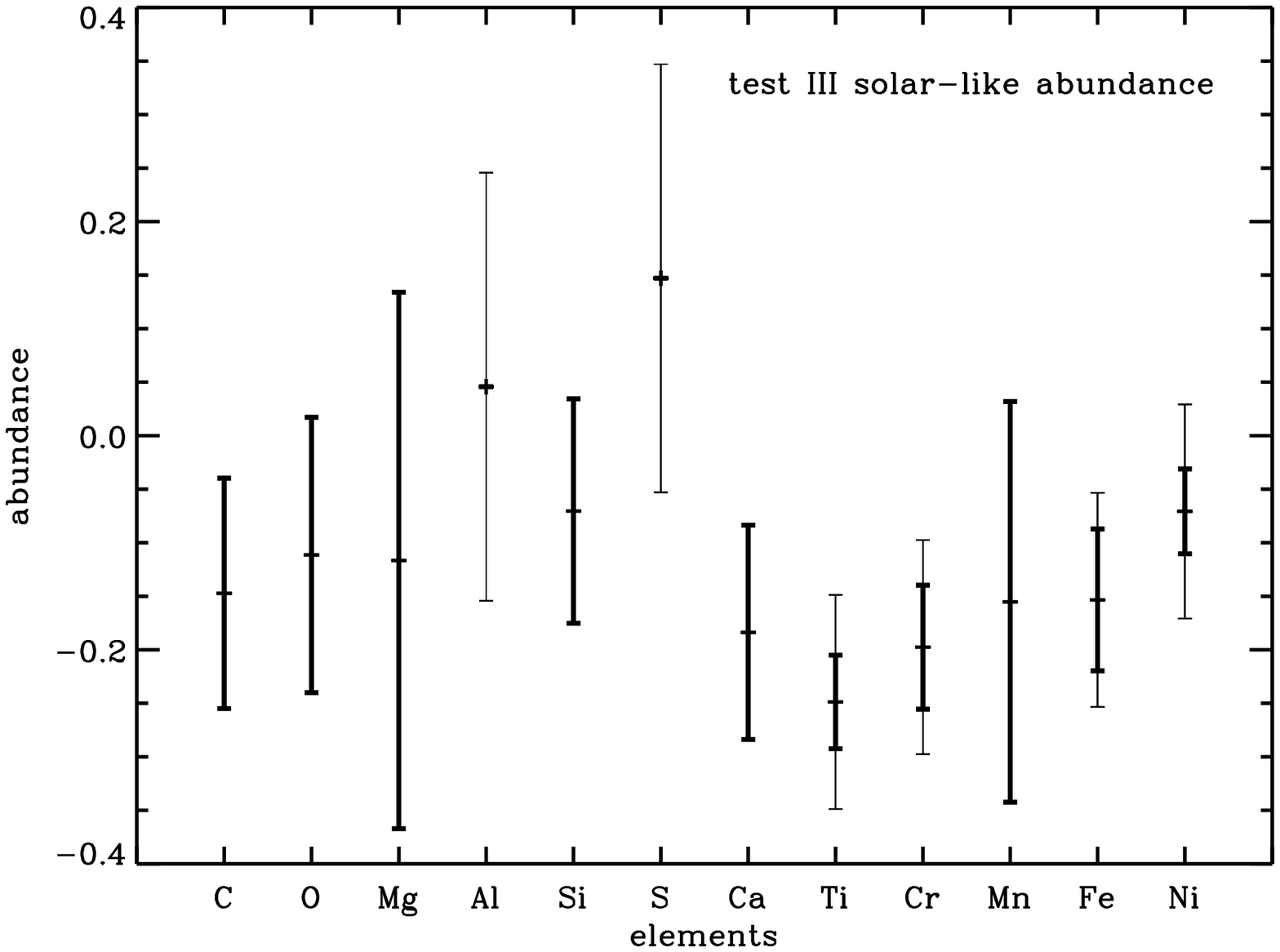}
\end{minipage}
\begin{minipage}{0.45\linewidth}
\centering
\includegraphics[width=\linewidth]{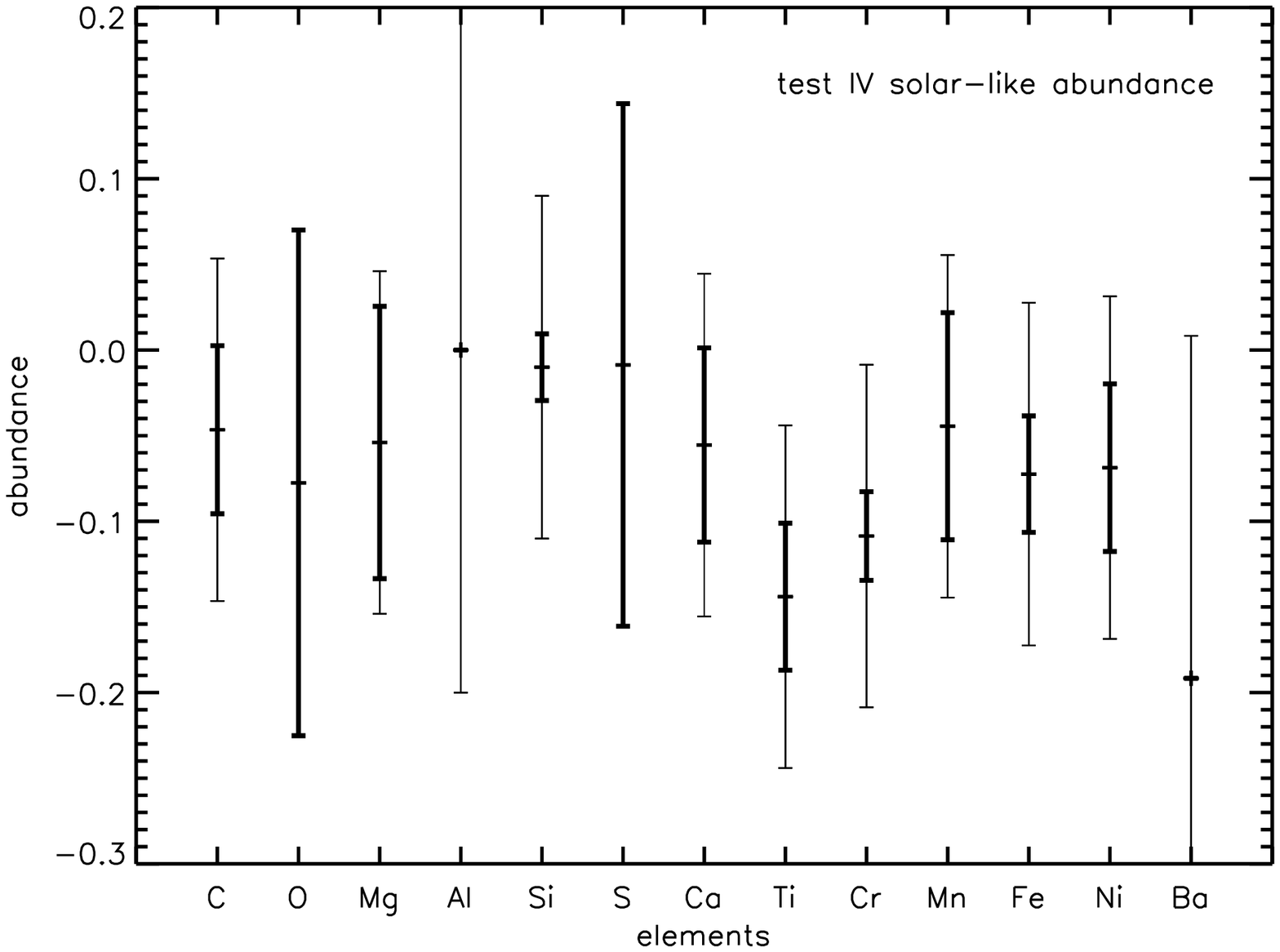}
\end{minipage}
\begin{minipage}{0.45\linewidth}
\centering
\includegraphics[width=\linewidth]{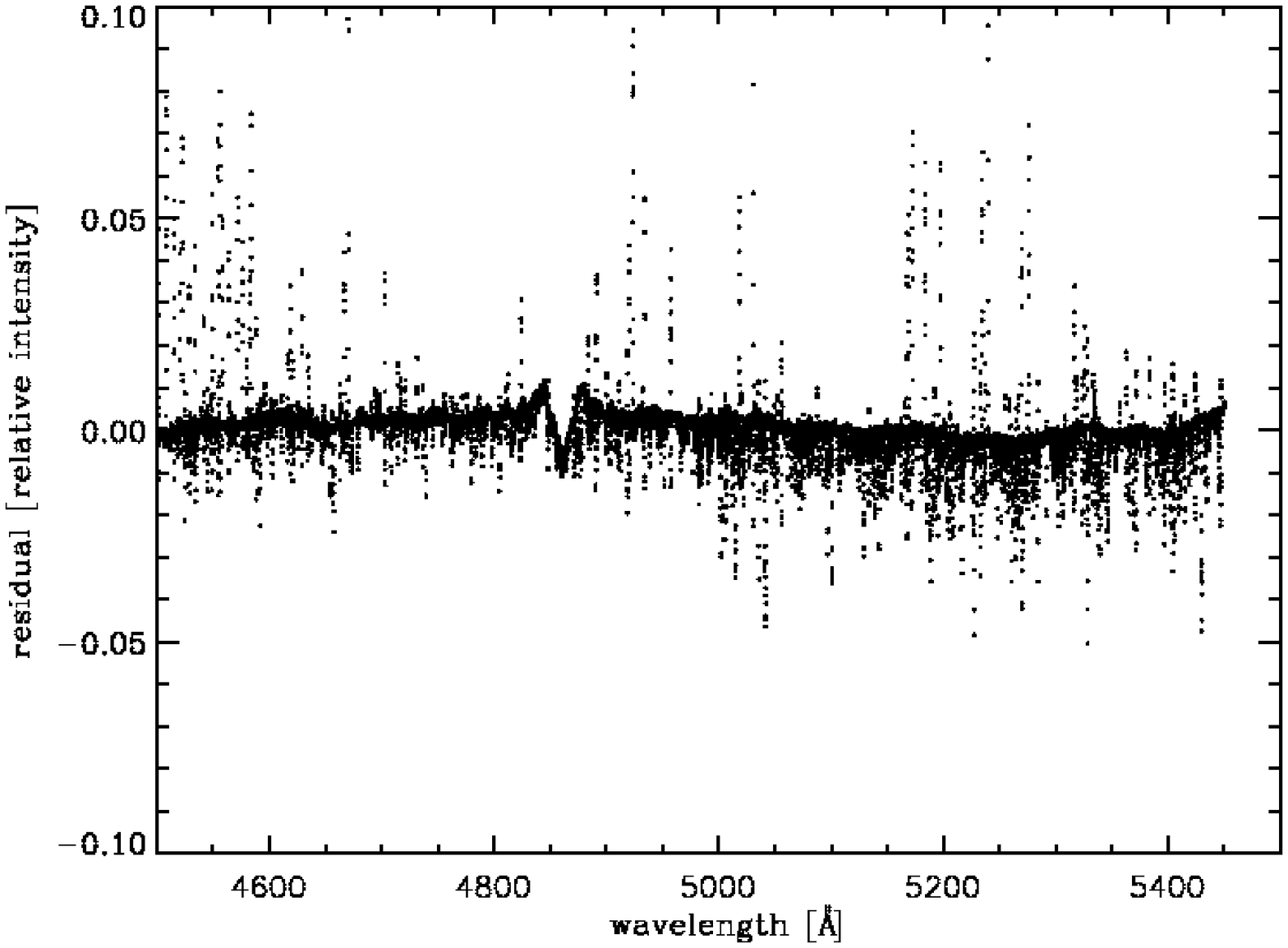}
\end{minipage}
\begin{minipage}{0.45\linewidth}
\centering
\includegraphics[width=\linewidth]{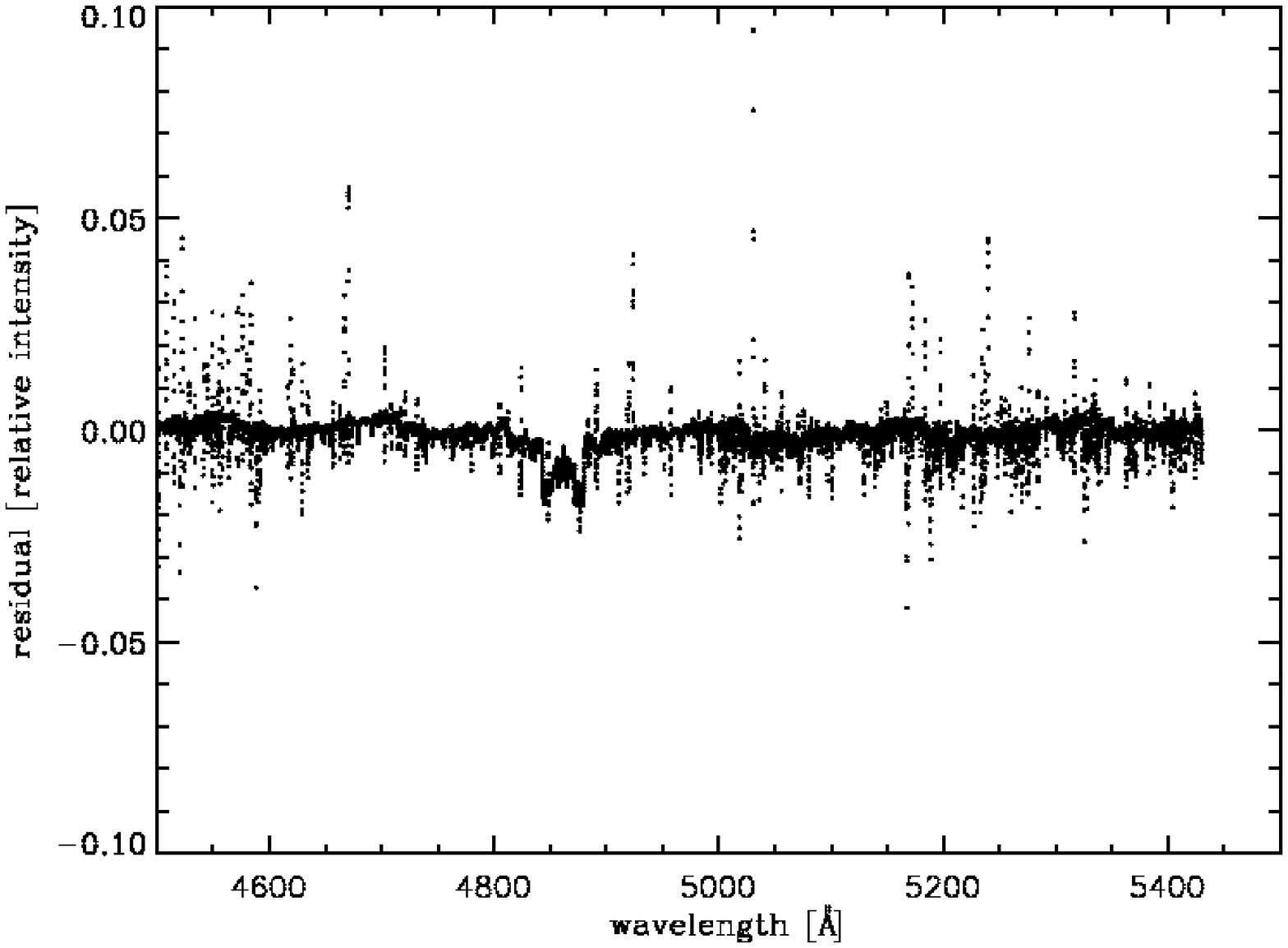}
\end{minipage}
\caption{{\sl Top:} Fitted abundances for star III (left) and star IV (right). The thick error bars are the standard deviations, which serve as the error when they are larger than 0.1. The thin error bars are the inferred errors of 0.1 for an element fitted at several spectral ranges and 0.2 for an element fitted in only one spectral region. {\sl Bottom:} Residuals of the input spectra minus the fitted spectra for the full wavelength region.}
\label{abunrottest}
\end{figure*}

Figure~\ref{abunvega} compares the abundances we obtained for the different elements to those from previous works, i.e., \citet{adelman1990}, \citet{qiu2001}, \citet{erspamer2002} and \citet{yoon2008}. The last one is the only analysis taking gravitational darkening due to rotation into account, hence the difference in values of the stellar parameters at the poles and equator. The scatter in abundances between the different analyses is considerable. These differences might originate from the inclusion of the deformation due to rotation or from the use of different stellar parameters: \citet{adelman1990} and \citet{erspamer2002} used ATLAS model atmospheres available for Vega \citep{kurucz1979} with T$_{\rm eff}$ = 9400 K, $\log$ g = 3.9 and a $\xi_{\rm micro}$ of 2 and 0 km\,s$^{-1}$, respectively, while \citet{qiu2001} used T$_{\rm eff}$ = 9430 K, $\log$ g = 3.95 and $\xi_{\rm micro}$ = 1.5 km\,s$^{-1}$, which are somewhat cooler than our model atmosphere. Furthermore, for some elements only a few lines can be fitted which make the final abundance values very sensitive to line parameters such as $\log$ gf. \citet{erspamer2002} argue that differences of some tenths of dex is not surprising for results from different authors, all using plane-parallel models. Our results are in good agreement with the abundances found by \citet{erspamer2003} who used data obtained with the same instrument.
Gravitional darkening due to fast rotation (such as in Vega) further introduces line profile deformations and biases on the chemical abundance determinations that are more or less pronounced \citep{yoon2008,takeda2008}, depending on the line formation region (e.g., equator, poles or the whole surface). The offsets seen in Fig.~\ref{abunvega} are consistent with their predictions.
We further discuss the effects of fast rotation on chemical abundance determination in the next section.

\section{Influence of fast rotation on present analyses \label{fastrot}}
Here, we investigate what offsets in abundance we would expect to find for fast-rotating stars seen nearly pole-on with our stellar parameter and abundance determination procedure. We simulated two spectra (star III and IV) with the same T$_{\rm eff}$ and $\log$ g at the poles and the same rotational velocity, but with different inclination angles. These simulations are based on model 4 of \citet{takeda2008}, and have solar abundance. The input and computed parameters for both simulated stars are given in Table~\ref{paramrottest}. 

The mean difference between the input and fitted spectrum is 0.0003 and -0.001 for star III and star IV respectively with standard deviations of 0.0075 and 0.0045. The large spikes in the residuals are due to line deformations. The large feature in the residuals at about 4860 \AA~is due to rotation and could serve as an indication of fast rotation when analysing real stars. The shape of the residual changes as a function of the difference in stellar parameter values at the pole and at the equator. Nevertheless, the values we obtained for T$_{\rm eff}$ and $\log$ g are within the range of the input values.

\begin{figure}
\begin{minipage}{\linewidth}
\centering
\includegraphics[width=\linewidth]{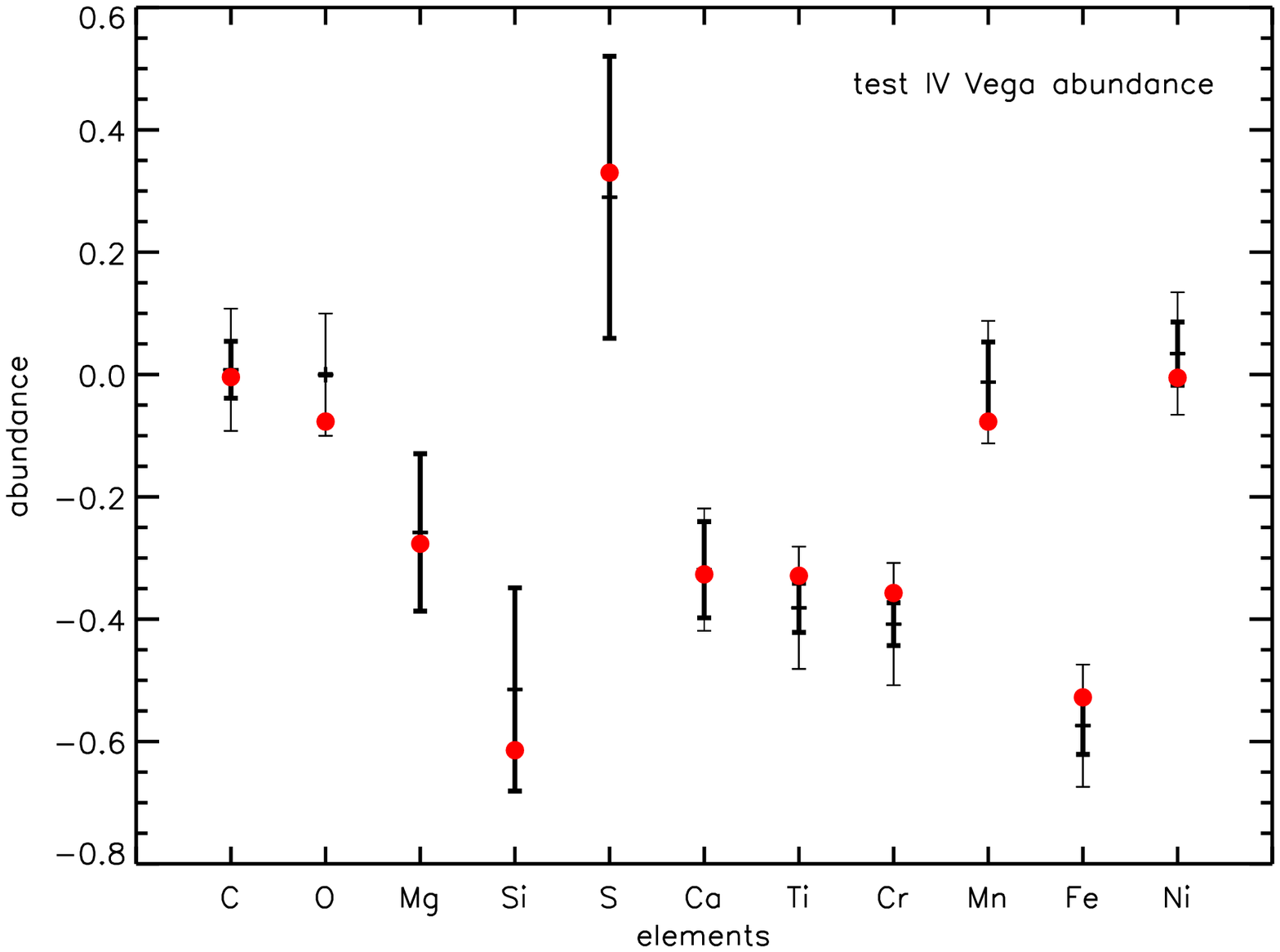}
\end{minipage}
\begin{minipage}{\linewidth}
\centering
\includegraphics[width=\linewidth]{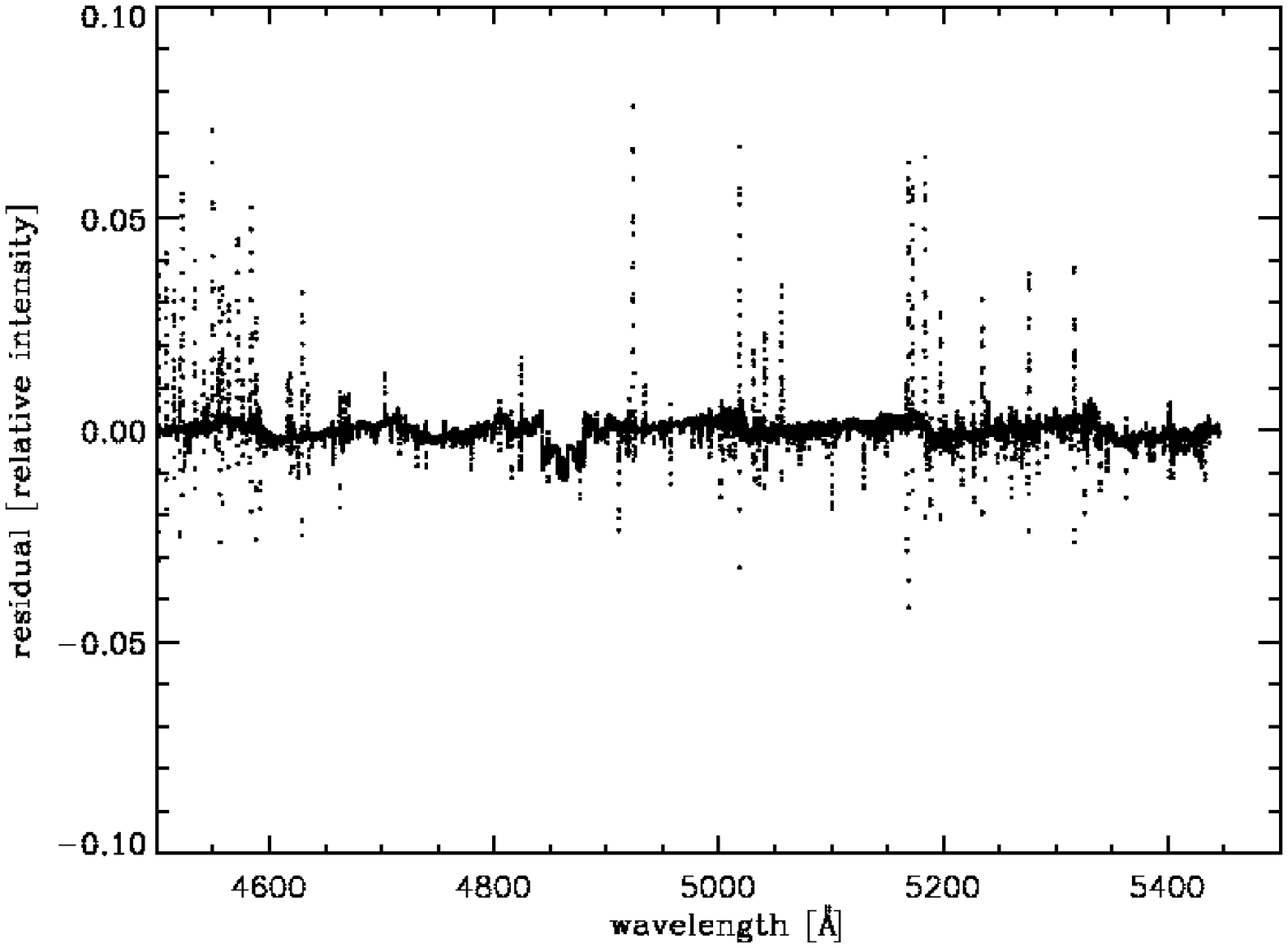}
\end{minipage}
\caption{{\sl Top:} Fitted abundances for star IV with input abundances for Vega, indicated with red dots (colours only in the online version). The thick error bars are the standard deviations, which serve as the error when they are larger than 0.1. The thin error bars are the inferred errors of 0.1 for an element fitted at several spectral ranges and 0.2 for an element fitted in only one spectral region. {\sl Bottom:} Residuals of the input spectra minus the fitted spectra for the full wavelength region.}
\label{abunrottestvega}
\end{figure}

Results of the chemical abundance analyses are shown in Fig.~\ref{abunrottest}. One notices that the abundances we derive are sensitive to the inclination angle, i.e., star III provides stronger deviations than star IV (both solar-like abundances). Gravitational darkening effects due to rotation often lead to apparent chemical underabundances in A0 stars. Star III indeed shows underabundances ranging from 0 to $-$0.25 for all considered elements, except for Al and S, which are found to be overabundant. It is interesting to note that we observe the same abundance pattern for Vega (Fig.~\ref{abunvega}), although more pronounced. Also, the residuals in the H$\beta$ line core are very similar for both Vega and star III.  So, if Vega were at an inclination angle of about 4$^{\circ}$, the gravitational darkening effects due to fast rotation could account for part of the chemical peculiarities of Vega.

For star IV we recover the input abundances without any significant offsets, which indicates that for stars with inclination angles $> 7^{\circ}$ we can recover the `real' abundances. According to \citet{takeda2008} the inclination angle of Vega is $\sim 7^{\circ}$, in which case we would be able to recover the `real' abundances for this star. Computations of star IV, with the abundances we obtained for Vega confirm this (see Fig.~\ref{abunrottestvega}).

\begin{figure*}
\begin{minipage}{0.45\linewidth}
\centering
\includegraphics[width=\linewidth]{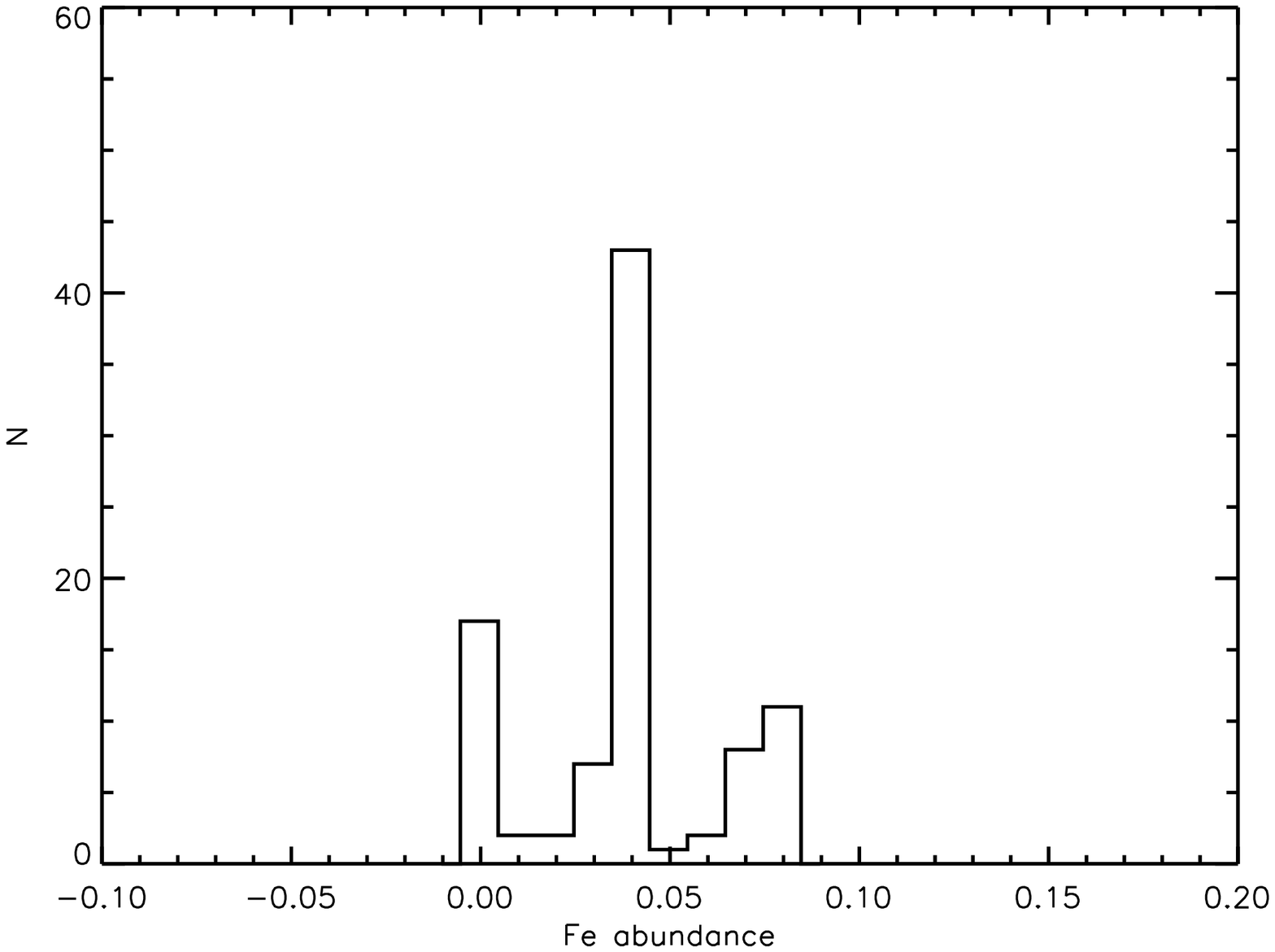}
\end{minipage}
\begin{minipage}{0.45\linewidth}
\centering
\includegraphics[width=\linewidth]{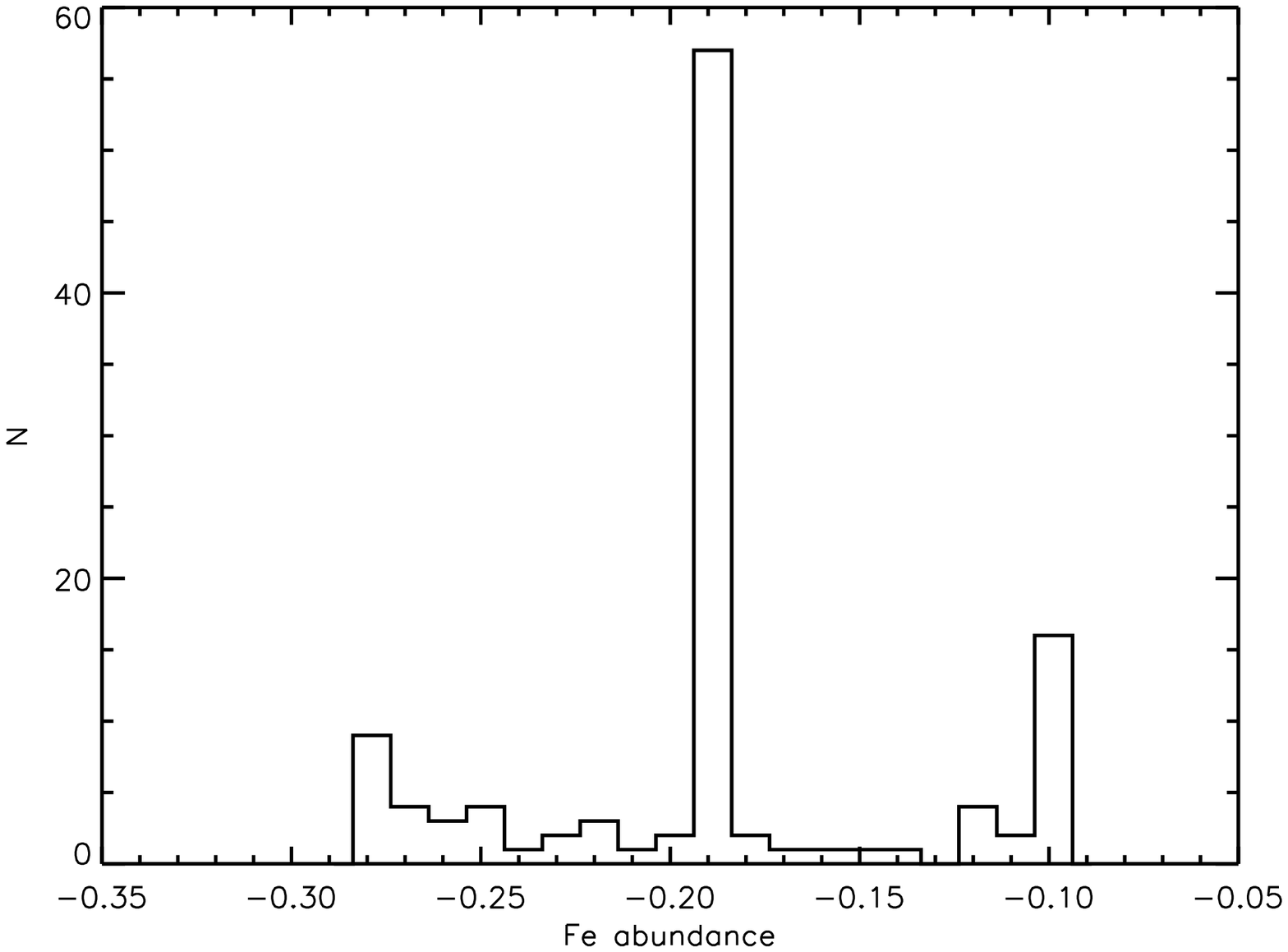}
\end{minipage}
\begin{minipage}{0.45\linewidth}
\centering
\includegraphics[width=\linewidth]{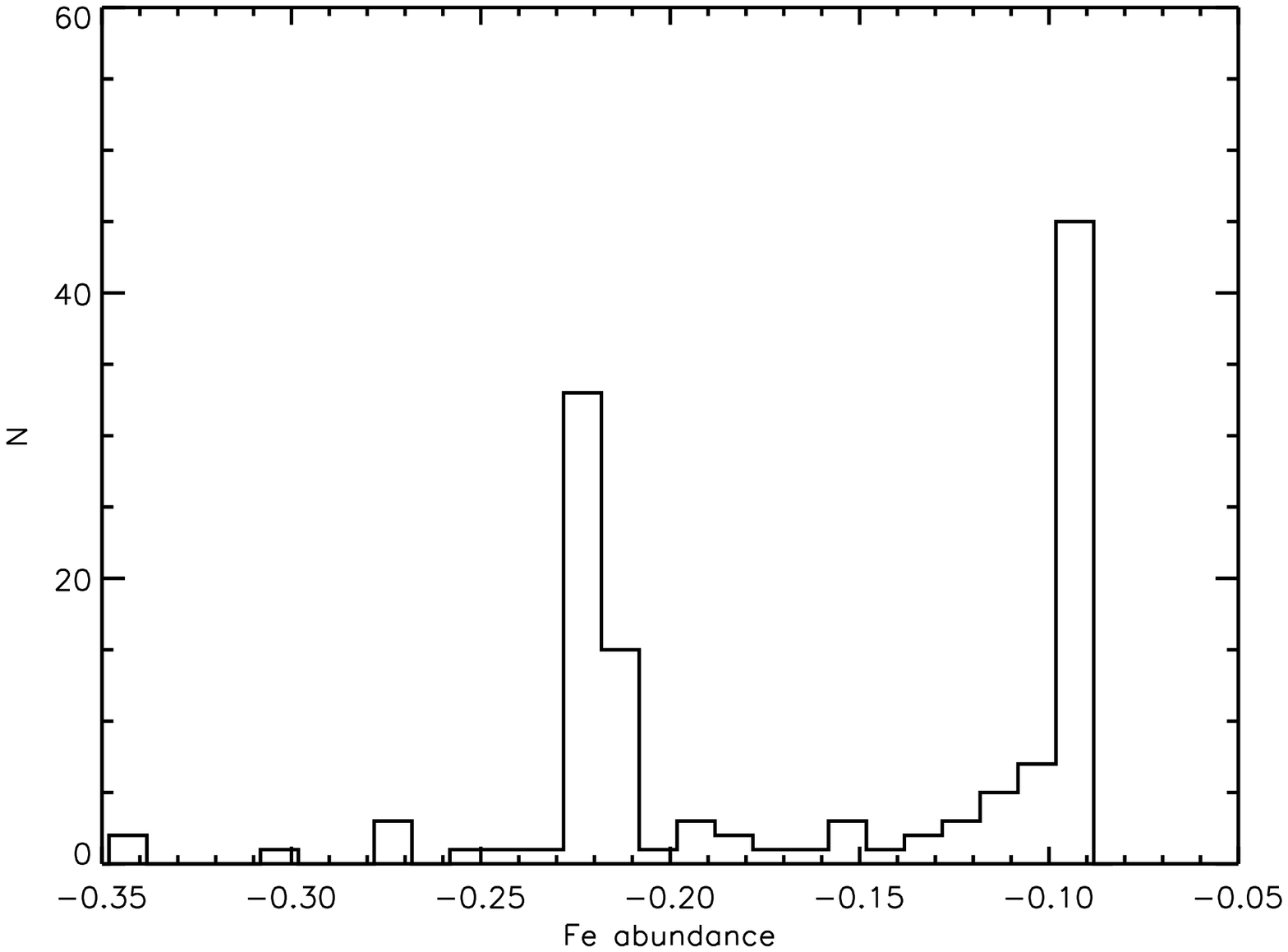}
\end{minipage}
\begin{minipage}{0.45\linewidth}
\centering
\includegraphics[width=\linewidth]{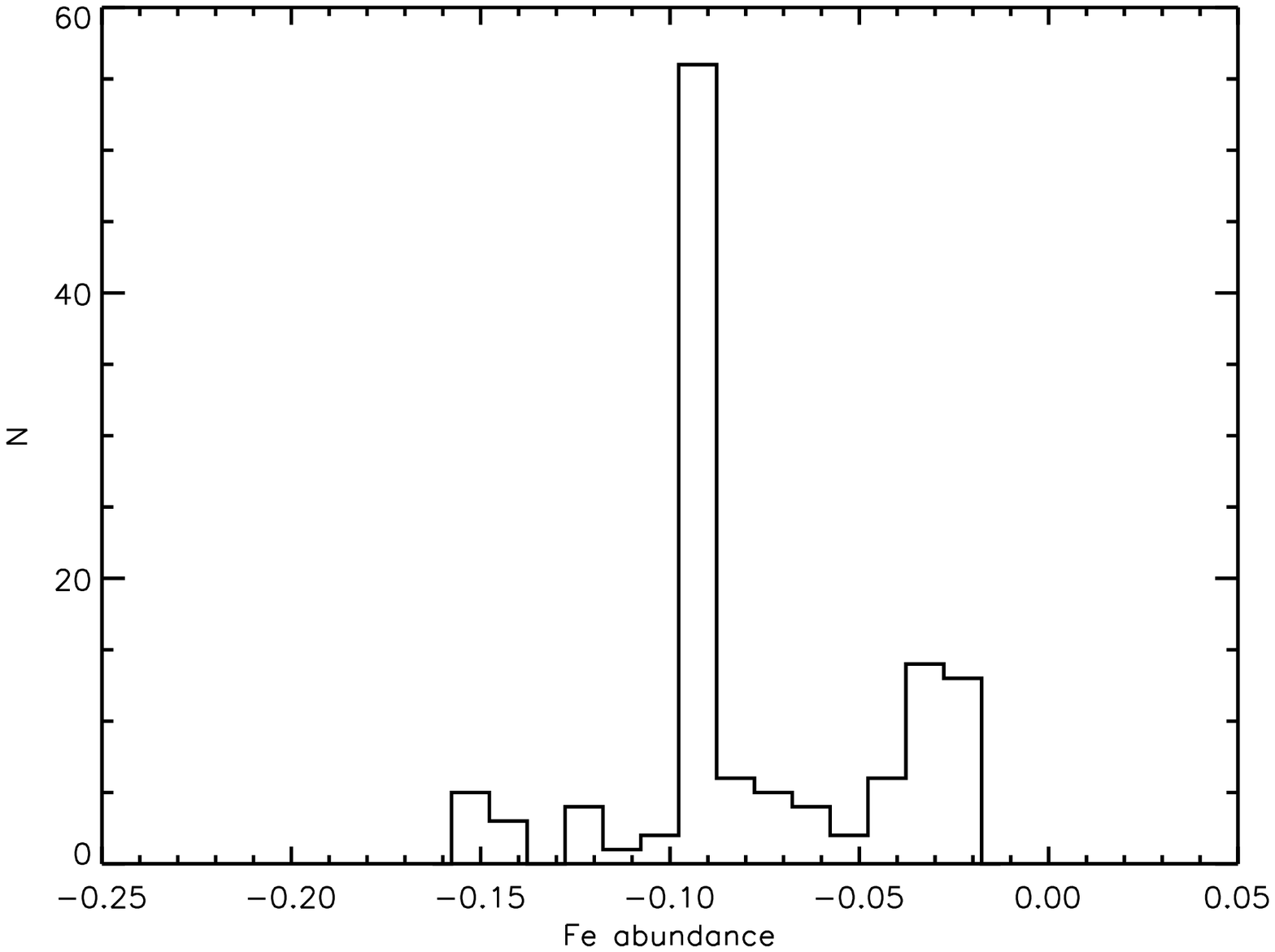}
\end{minipage}
\caption{Distributions of iron abundance obtained for the different regions of the spectrum for star I (left top), star II (right top), star III (left bottom) and star IV (right bottom).}
\label{Fedist}
\end{figure*}

\citet{yoon2008} used a model for Vega with an inclination angle of 4.54$^{\circ}$, which is closer to our test star III. They mention that assigning a single abundance and $\xi_{micro}$ for all lines of an element with low ionisation and excitation potentials, such as Fe I, is inadequate, because these elements are dominated by the contribution from the equatorial region causing a double-horned shape in the spectral lines. A single abundance value for each element is however assumed in our procedure. To further investigate this, we study the distribution of the iron abundances fitted for different spectral regions. The distributions for all four test stars are shown in Fig.~\ref{Fedist}. Clearly, star III with the lowest inclination angle has a bimodal distribution and thus the average of the abundance is incorrect for most lines, confirming the findings of \citet{yoon2008}. 
The bimodal effect is already much less pronounced in star IV with an inclination angle of 7.4$^{\circ}$. The distributions of the slowly rotating stars I and II are more symmetric and better centred. 

\begin{table*}
\centering
\begin{minipage}{\linewidth}
\centering
\caption{Stellar parameters of the two test spectra including fast-rotation. The errors indicated for the computed values are the standard deviations of the values determined in different wavelength regions. $\Omega/\Omega_c$ is the ratio of the rotational velocity to the break-up velocity. }
\label{paramrottest}
\begin{tabular}{lrrrr}
\hline
 & star III & star III & star IV & star IV\\
 & input & computed & input & computed\\
 \hline
T$_{\rm eff}$ (pole) [K] & 9826 & - & 9829 & -\\
$\log$ g (pole) (c.g.s) & 3.97 & - & 3.97  & -\\
T$_{\rm eff}$ (equator) [K] & 7971 & - & 9408 & -\\
$\log$ g (equator) [cm\,s$^{-2}$] & 3.61 & - & 3.89  & -\\
T$_{\rm eff}$  [K] & - & 9133 $\pm$ 117 & - & 9677 $\pm$ 66\\
$\log$ g  (c.g.s) & - & 3.80 $\pm$ 0.02 & - & 3.95 $\pm$ 0.05\\
$v \sin i$ [km\,s$^{-1}$] & 15.7 & 13.2 $\pm$ 2.4 & 15.7 & 13.6 $\pm$ 1.7\\ 
inclination [deg] & 3.8 & - & 7.39 & -\\
$\Omega/\Omega_c$ & 0.8 & - & 0.44 & -\\
$\xi_{\rm micro}$ [km\,s$^{-1}$] &  2.0 & 2.9 $\pm$ 1.3 & 2.0 & 2.25 $\pm$ 1.2 \\
\hline
\end{tabular}
\end{minipage}
\end{table*}

\section{Summary and Conclusions \label{sumcon}}
In this work, we presented a semi-automatic procedure developed to determine stellar parameters and abundances of elements from helium to mercury for A- and F-type stars. We performed various tests with this stand-alone procedure on synthetic spectra which had a resolving power of 40\,000 and a wavelength range of 4000-7000 \AA.

Most abundance analyses are applied to a well-observed object like the Sun to verify and demonstrate its performance and dependability. In this study we investigate the accuracy of our analyses on synthetic test spectra. The main advantage of this is that we can perform the abundance analysis with correct stellar parameters, i.e., the known T$_{\rm eff}$, $\log$ g and $\upsilon \sin i$ for which the test spectrum is computed and use similar $\log$ gf values for the "observation" as well as for the computations. This approach allows us to test the influence of erroneous stellar parameters on the abundance determination. From the analyses of these test spectra, we can conclude that we are able to obtain abundances which are consistent with the input values,  within the errors. When "reasonable" changes (i.e. changes due to reasonable observational errors) of the stellar parameters are considered, we show that the variance in the resulting abundances is smaller than the difference between the input and the computed values. From this, we conclude that stellar parameters which have a reasonable offset from their `real' values have only minor influence on the abundance determination.

Our method is also tested on an observed spectrum of Vega, which is a fast rotator seen nearly pole-on, and compared with previous studies. The published results for Vega show that there are rather large differences between the abundances from different studies. Our results are in good agreement with the abundances found by \citet{erspamer2002} who used data obtained with the same instrument.

Although we see only a small fraction of fast-rotating stars nearly pole-on, we might encounter one or more of them. We investigated this effect by simulating spectra which include fast rotation seen nearly pole-on.
The computed abundances for the test star with an inclination angle of $\approx 4 ^{\circ}$ are offset from the input values, while this offset is negligible at an inclination angle of 7$^{\circ}$. This is also reflected in the bimodal distribution of the iron abundances obtained in different regions of the spectrum. This bimodal behaviour is likely due to the fact that lines with lower excitation potentials are more influenced by the rapidly rotating equatorial regions than lines with higher excitation potentials. For stars with an inclination angle of $\ge 7^{\circ}$ the offsets from the input values are negligible and also the distribution of iron abundances obtained for different spectral regions is almost single-peaked.

From our simulations we conclude that we should be able to recover the `real' abundance values of Vega, if this star is indeed observed with an inclination angle of about 7$^{\circ}$, as suggested by \citet{takeda2008}. However, if Vega has an inclination angle of about 4$^{\circ}$, as used by \citet{yoon2008}, the influence of rotation should indeed be visible in the computed abundances. Note that the residual shape of H$\beta$ of Vega is more similar to our simulations with the lowest inclination angle.

The procedure for stellar parameter determination and abundance analysis, described in Sections 2 and 3, and tested on synthetic spectra and Vega, will now be applied to observations of A and F stars that we have at our disposal.  This sample of stars contains both single stars and multiple systems, and these will allow us to study the relationship between pulsations, multiplicity and chemical composition.  We have also obtained spectra of suspected $\gamma$ Dor stars, with the aim of both confirming their classification and applying this abundance procedure.  Results will be presented in subsequent publications.

\section*{Acknowledgments}
SH, YF, PL and PDC acknowledge financial support from the Belgian Federal Science Policy (ref: MO/33/018).  We acknowledge funding by the Optical Infrared Co-ordination network (OPTICON), a major international collaboration supported by the Research Infrastructures Programme of the European Commission's Sixth Framework Programme. This research was in part supported by the European Helio- and Asteroseismology Network (HELAS), a major international collaboration funded by the European Commission's Sixth Framework Programme. We would like to thank the anonymous referee for valuable comments, which helped to improved the manuscript considerably.


\label{lastpage}

\end{document}